\documentclass[twocolumn,10pt]{asme2ej}

\usepackage{epsfig} 

\title{Real-time Eco-Driving Control in Electrified Connected and Autonomous Vehicles using Approximate Dynamic Programming}

\author{Shreshta~Rajakumar~Deshpande \thanks{Address all correspondence to this author.}
    \affiliation{
	Center for Automotive Research\\
	Department of Mechanical Engineering\\
	The Ohio State University\\
	Columbus, OH, USA - 43212\\
    Email:  rajakumardeshpande.1@osu.edu
    }
}

\author{Shobhit~Gupta \\
	Center for Automotive Research\\
	Department of Mechanical Engineering\\
	The Ohio State University\\
	Columbus, OH, USA - 43212\\
	Email: gupta.852@osu.edu
}

\author{Abhishek~Gupta \\
	Department of Electrical and Computer Engineering\\
	The Ohio State University\\
	Columbus, OH, USA - 43210\\
	Email: gupta.706@osu.edu
}

\author{
	Marcello~Canova \\
	Center for Automotive Research\\
	Department of Mechanical Engineering\\
	The Ohio State University\\
	Columbus, OH, USA - 43212\\
	Email: canova.1@osu.edu
}

\usepackage{tikz}
\usepackage{verbatim}
\usetikzlibrary{arrows,shadows,shapes,fit,scopes,calc,matrix,positioning,decorations.pathmorphing}

\usepackage{graphicx}
\usepackage{multirow, array}
\usepackage{subfig, caption, float}
\usepackage{color,soul}
\usepackage{amsmath, mathrsfs, mathtools, amsfonts}
\DeclarePairedDelimiter\abs{\lvert}{\rvert}%
\usepackage{hyperref}
\usepackage{siunitx}
\usepackage{dblfloatfix}
\usepackage{xcolor}

\begin{document}

\maketitle    

\begin{abstract}
{\it Connected and Automated Vehicles (CAVs), particularly those with a hybrid electric powertrain, have the potential to significantly improve vehicle energy savings in real-world driving conditions. In particular, the Eco-Driving problem seeks to design optimal speed and power usage profiles based on available information from connectivity and advanced mapping features to minimize the fuel consumption over an itinerary. This paper presents a hierarchical multi-layer Model Predictive Control (MPC) approach for improving the fuel economy of a 48V mild-hybrid powertrain in a connected vehicle environment. Approximate Dynamic Programming (ADP) is used to solve the Receding Horizon Optimal Control Problem (RHOCP), where the terminal cost for the RHOCP is approximated as the base-policy obtained from the long-term optimization.

The controller was extensively tested virtually (using both deterministic and Monte Carlo simulations) across multiple real-world routes where energy savings of more than 20\%  have been demonstrated. Further, the developed controller was deployed and tested at a proving ground in real-time on a test vehicle equipped with a rapid prototyping embedded controller. Real-time in-vehicle testing confirmed the energy savings observed in simulation and demonstrated the ability of the developed controller to be effective in real-time applications.
}
\end{abstract}


\section{Introduction}
\label{sec::introduction}


\underline{C}onnected and \underline{A}utomated \underline{V}ehicles (CAVs) have the potential to increase safety, driving comfort, as well as fuel economy, by exploiting look-ahead driving information available via advanced navigation systems, vehicle-to-vehicle (V2V) and vehicle-to-infrastructure (V2I) communication \cite{guanetti2018control,gupta2020estimation}. Intuitively, with these connectivity technologies, a controller can plan a speed trajectory that minimizes unnecessary acceleration and braking events, thereby improving driver comfort and fuel-efficiency \cite{xu2018design}. Meanwhile, powertrain electrification can increase the vehicle fuel (or energy) efficiency, by including battery packs and electric motors as alternative energy storage and power generation devices respectively \cite{guzzella2007vehicle}. While combining these two technologies could further compound their efficiency improvements, they present a greater challenge from a planning and control perspective \cite{alam2014critical}.

Eco-Driving, defined in literature as the control of vehicle velocity for minimizing the fuel or energy consumption over an itinerary, is gaining prominence \cite{sciarretta2015optimal}. Based on the powertrain configuration considered in each application, the methods developed for determining the optimized velocity profile can vary, depending on whether the powertrain is equipped with a single power source \cite{jin2016power, ozatay2014cloud, han2019fundamentals} or a hybrid electric drivetrain \cite{mensing2012vehicle, guo2016optimal}. The latter involves modeling multiple power sources and devising optimal control algorithms that can split the power demand in a synergistic manner to efficiently utilize the electric energy stored in the battery. Most of the existing literature consider the Eco-Driving problem in a decentralized fashion i.e, the speed trajectory optimization and powertrain control are performed sequentially, and then integrated within a hierarchical control structure \cite{amini2019sequential}. Based on overall system efficiency and calibration effort considerations, this paper explores an alternative approach, where the speed trajectory and powertrain torque split are jointly optimized.

Incorporation of V2I information enables the processing and use of \underline{S}ignal \underline{P}hase \underline{a}nd \underline{T}iming (SPaT) information in real-time for performing \underline{Eco}-\underline{A}pproach a\underline{n}d \underline{D}eparture (Eco-AND) at signalized intersections. Eco-AND refers to velocity control for increasing the likelihood to pass through a signalized intersection during a green window. Studies from literature have shown that such maneuvers can improve the overall fuel efficiency \cite{hao2018eco,ye2018prediction,altan2017glidepath}. When Eco-AND is combined with Eco-Driving, the added complexity means that the controller may not always be able to compute a speed trajectory without violating traffic rules (i.e., optimization constraints) \cite{sun2018robust}. This warrants a unified optimization framework combining Eco-Driving and Eco-AND that can improve the fuel efficiency and ensure constraint satisfaction in a broad range of driving scenarios, while remaining computationally tractable on embedded control hardware platforms.


This work focuses on the development of a hierarchical multi-layer Eco-Driving controller based on \underline{M}odel \underline{P}redictive \underline{C}ontrol (MPC), that jointly optimizes the vehicle speed and battery \underline{S}tate \underline{o}f \underline{C}harge (SoC). The proposed control framework is a novel alternative to state-of-the-art, where the problem of velocity optimization and energy management is often solved in a decentralized fashion \cite{amini2019sequential}. To account for route uncertainty and limited V2I communication range, and to mitigate the intensive on-board computation, the optimization routine is split into a long-term and short-term optimal control problem as per the availability of the look-ahead information. The long-term optimization is performed over the entire route with the look-ahead information such as speed limits and position of stop-signs available from an advanced navigation system. To account for variability in route conditions such as variable speed limits and SPaT information, the full-route optimization is cast as a \underline{r}eceding \underline{h}orizon \underline{o}ptimal \underline{c}ontrol \underline{p}roblem (RHOCP) and evaluated periodically in real-time via MPC over a short horizon. While MPC has been widely used in real-time control applications \cite{qin2003survey}, a key challenge still remains in its implementation -- the definition of an appropriate terminal cost and/or terminal state constraints for performance and stability \cite{borrelli2017predictive}. Using principles from \underline{A}pproximate \underline{D}ynamic \underline{P}rogramming (ADP), specifically the rollout algorithm, this paper provides a novel methodology for appropriately selecting the terminal cost of the MPC.

The proposed solution method yields optimal closed-loop strategies that are robust against external disturbances and has its theoretical foundations in the traditional \underline{D}ynamic \underline{P}rogramming (DP) \cite{bellman1966dynamic,lee2010approximate,yin2016energy}. This hierarchical optimization framework is well-suited for in-vehicle implementation, where advanced mapping systems can provide long-term route information and vehicle-to-everything (V2X) technologies can update this information in the short-term. Further, this work proposes a systematic and comprehensive verification framework for virtual and experimental evaluation of fuel-savings from the CAV optimization algorithms. Simulations and track testing results validate the developed rollout algorithm-based optimization framework making it suitable for real-time applications containing environmental disturbances and modeling uncertainties.

\section{Model of Parallel Mild-Hybrid Electric Vehicle}
\label{sec::model}

\begin{figure}[!b]
	\centering
	\includegraphics[width=0.9\columnwidth]{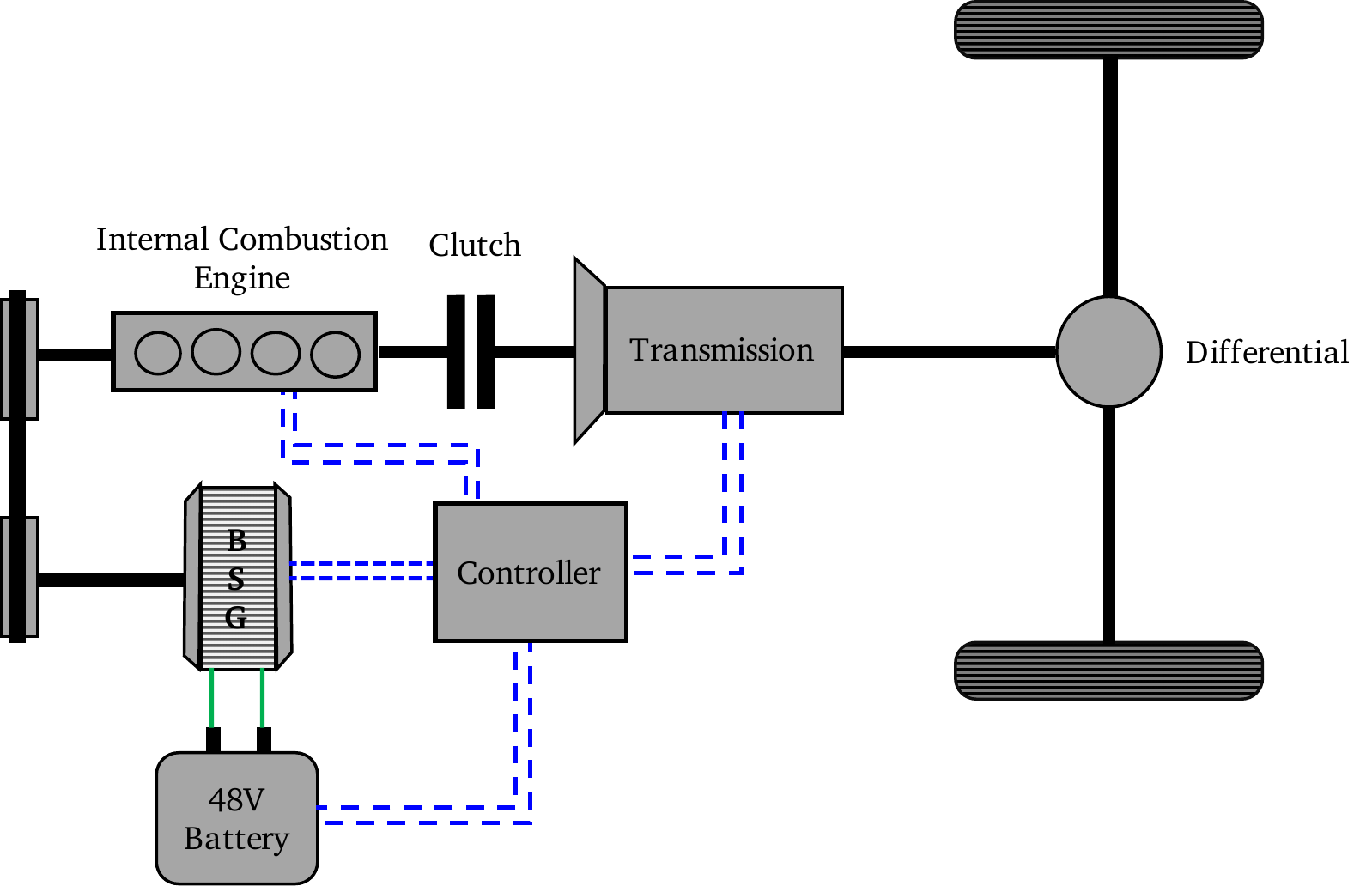}
	\caption{Block diagram of P0 mHEV topology.}
	\label{fig::NEXTCAR_mHEV_topology}
\end{figure}

The P0 \underline{m}ild-\underline{h}ybrid \underline{e}lectric \underline{v}ehicle (mHEV) architecture studied in this work is illustrated in Fig. \ref{fig::NEXTCAR_mHEV_topology}. A \underline{B}elted \underline{S}tarter \underline{G}enerator (BSG) is connected to the crankshaft of a $\SI{1.8}{L}$ turbocharged gasoline engine equipped with Dynamic Skip Fire, DSF \cite{wilcutts2013design, wilcutts2018electrified} and a $\SI{48}{V}$ battery pack.

\begin{figure*}[!t]
	\centering
	\includegraphics[width=1.4\columnwidth]{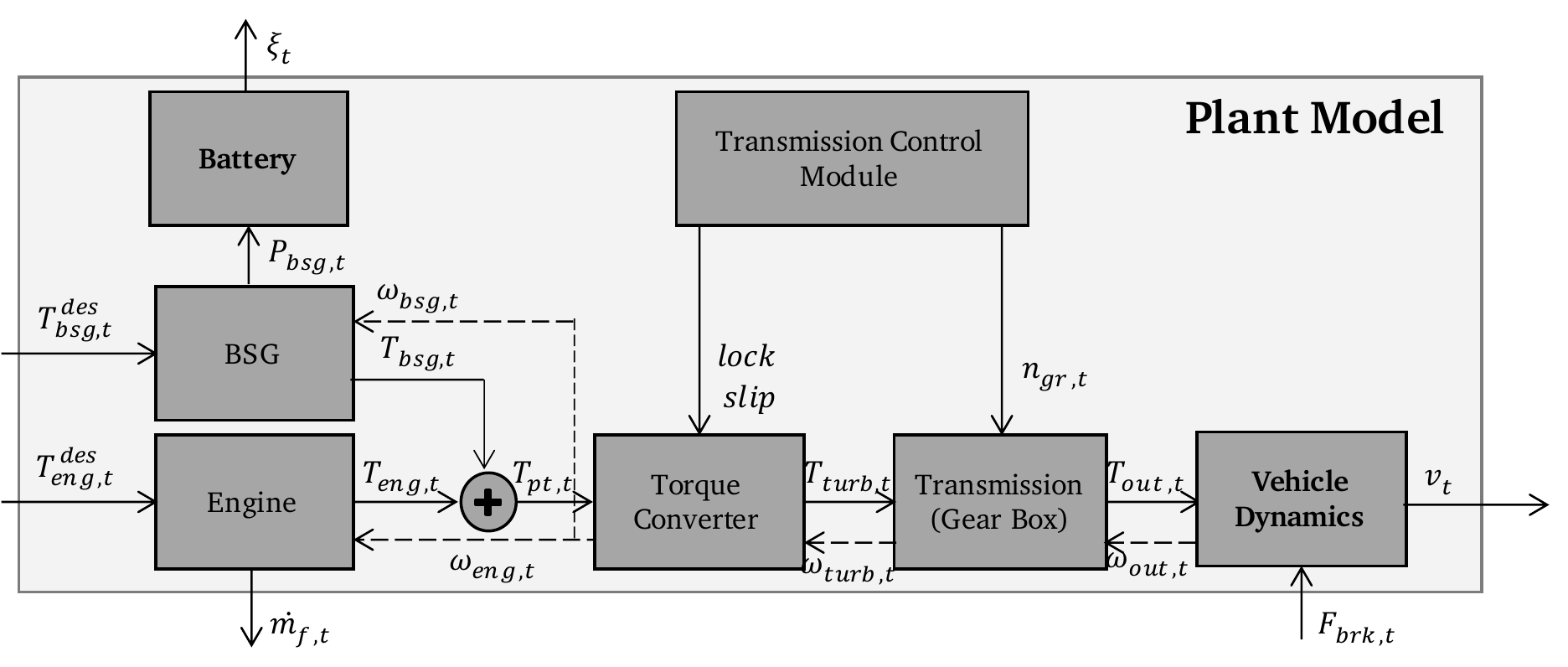}
	\caption{Block diagram of 48V P0 mild-hybrid drivetrain.}
	\label{fig::plant_model}
\end{figure*}

A forward-looking vehicle simulator was developed in MATLAB-Simulink to evaluate fuel consumption and compare different control strategies over prescribed routes. The inputs to the plant model in Fig. \ref{fig::plant_model} are the desired BSG torque ($T_{bsg,t}^{des}$) and desired engine torque ($T_{eng,t}^{des}$), which are obtained from a simplified model of the software in the \underline{E}lectronic \underline{C}ontrol \underline{M}odule (ECM). This contains a baseline torque split strategy and essential functions that convert the driver's input (pedal positions) to commands, which are fed to the powertrain components. Note that in the descriptions of the equations that follow, the arguments of certain terms may be suppressed for brevity.

\subsection{Vehicle Dynamics}


The vehicle acceleration is modeled with the road-load equation, considering only the longitudinal dynamics and disregarding the lateral dynamics \cite{guzzella2007vehicle}:
\begin{equation}
\label{eq::longitudinal_dyn_model}
\begin{aligned}
\frac{\mathrm{d}v_t}{\mathrm{d}t} &= \frac{F_{trc,t} - F_{road,t}(v_t)}{M} \\
F_{trc,t} &= \frac{T_{out,t}}{R_w}-F_{brk,t}
\end{aligned}
\end{equation}
where $v_{t}$ is the velocity of the vehicle, $M$ is the effective mass of the vehicle (including driveline rotational inertia and the total payload), $F_{trc,t}$ is the net force exerted by the propulsion system (including the braking force $F_{brk,t}$) of the vehicle, $R_w$ is the rolling radius of the wheel, and $F_{road,t}$ is the road-load, which is defined as the force imparted on a vehicle from aerodynamic drag, tire rolling resistance and road grade:
\begin{equation}
\label{eq::road_load}
\begin{aligned}
F_{road,t}\left(v_{t} \right) =  \dfrac{1}{2}C_d\rho_a A_f v_{t}^2 &+ Mg \cos{\alpha} \cdot C_{r,t}\left(v_{t} \right) \\
& + Mg\sin{\alpha}
\end{aligned}
\end{equation}
where $C_d$ is the aerodynamic drag coefficient, $\rho_a$ is the air density, $A_f$ is the effective aerodynamic frontal area, $C_{r,t}$ is rolling resistance coefficient, and $\alpha$ is the road grade (expressed in $\si{rad}$).

\subsection{Powertrain}
The engine fuel consumption is modeled using static nonlinear maps $\psi_t (\cdot,\cdot,\cdot)$, as functions of the gear, engine speed and torque. The fuel flow rate obtained from this model can be expressed as:
\begin{equation}
\label{eq::fuel_consumption}
\begin{aligned}
\dot{m}_{f,t} = \psi_t(n_{gr,t}(v_{t},T_{eng,t}),\omega_{eng,t},T_{eng,t})
\end{aligned}
\end{equation}
where $n_{gr,t}$ is the selected gear number, $T_{eng,t}$ is the engine torque, and $\omega_{eng,t}$ is the engine speed.


The BSG is modeled using an efficiency map, which is a quasi-static representation of torque production:
\begin{equation}
\label{eq::BSG_model}
\begin{aligned}
\omega_{bsg,t} &= r_{belt}\cdot \omega_{eng,t}\\
P_{bsg,t} &= T_{bsg,t} \cdot \omega_{bsg,t} \cdot \bar{\eta}_{bsg,t} \\
\bar{\eta}_{bsg,t} &= {\begin{cases}
	\eta_{bsg,t}(\omega_{bsg,t},T_{bsg,t}), & T_{bsg,t} < 0 \\
	\frac{1}{\eta_{bsg,t}(\omega_{bsg,t},T_{bsg,t}) }, & T_{bsg,t} > 0
	\end{cases}}
\end{aligned}
\end{equation}
where $\omega_{bsg,t}$ is the BSG speed, $r_{belt}$ is the belt ratio, $P_{bsg,t}$ is the electrical power required to produce a torque $T_{bsg,t}$ at speed $\omega_{bsg,t}$, and $\eta_{bsg,t}$ is the BSG efficiency.


The battery is modeled as a zero-th order equivalent circuit, which comprises an ideal voltage source and a resistor in series. The voltage across the circuit is considered the terminal battery voltage. The model equations are:
\begin{equation}
\label{eq::battery_model}
\begin{aligned}
I_{batt,t} &= \frac{V_{oc,t}(\xi_t) - \sqrt{V^2_{oc,t}(\xi_t) -4R_0\cdot P_{bsg,t}}}{2R_0} \\
\bar{I}_{batt,t} &= I_{batt,t} + I_{bias} \\
\frac{\mathrm{d}\xi_t}{\mathrm{d}t} &= -\frac{1}{C_{nom}}\cdot \bar{I}_{batt,t}
\end{aligned}
\end{equation}
where $V_{oc,t}$ is the battery open-circuit voltage, $R_0$ is an approximation of the battery internal resistance, $I_{batt,t}$ is the battery current, $\xi_t$ is the battery SoC, and $C_{nom}$ is the nominal capacity of the battery. Further, a calibration term $I_{bias}$ is introduced as a highly simplified representation of the on-board electrical auxiliary loads.

Power at the flywheel is transmitted to the wheels via a torque converter and $6$-speed automatic gearbox. A simplified torque converter model is developed with the purpose of computing the power losses during DSF activity. The desired clutch slip, $\omega_{slip,t}^{des}$, is determined by the transmission control module using maps as a function of the gear, engine speed and torque. The model equations are:
\begin{equation}
\label{eq::TC_model}
\begin{aligned}
\omega_{p,t} &= \omega_{turb,t} + \omega_{slip,t}^{des}(n_{gr,t}(v_t,T_{eng,t}),\omega_{eng,t},T_{eng,t}) \\
\omega_{eng,t} &= \begin{cases}
\omega_{p,t}, & \omega_{p,t} \geq \omega_{eng,stall} \\
\omega_{idle}, & 0\leq \omega_{p,t} < \omega_{eng,stall} \\
0, & 0\leq \omega_{p,t} < \omega_{eng,stall} , stop = 1
\end{cases}\\
T_{turb,t} &= T_{pt,t}
\end{aligned}
\end{equation}
where $\omega_{p,t}$ is the speed of the torque converter pump, $\omega_{turb,t}$ is the speed of the turbine, $\omega_{eng,stall}$ is the speed at which the engine stalls, $\omega_{idle}$ is the idle speed (target) of the engine, $stop$ is a flag from the ECM indicating engine shut-off when the vehicle is stationary, $T_{turb,t}$ is the turbine torque, and $T_{pt,t}$ is the powertrain torque.

The transmission is modeled as a static gearbox with efficiency $\eta_{tran,t}$, which is determined empirically from vehicle test data:
\begin{equation}
\label{eq::transmission_model}
\begin{aligned}
\omega_{turb,t} &= r_{f} \cdot r_{gr,t}(n_{gr,t}(v_{t},T_{eng,t})) \cdot \frac{v_{t}}{R_w} \\
T_{out,t} &= r_{f} \cdot r_{gr,t}(n_{gr,t}(v_{t},T_{eng,t})) \cdot T_{turb,t} \cdot \bar{\eta}_{tran,t}\\
\bar{\eta}_{tran,t} &= \begin{cases}
\eta_{tran,t}(\omega_{turb,t},T_{turb,t}), & T_{turb,t} \geq 0 \\
\dfrac{1}{\eta_{tran,t}(\omega_{turb,t},T_{turb,t})}, & T_{turb,t} < 0
\end{cases}
\end{aligned}
\end{equation}
where $r_{f}$ is the final drive ratio, $r_{gr,t}$ is the gear ratio, and $T_{out,t}$ is the transmission output shaft torque.

\begin{figure}[!t]
	\centering
	\vspace{-3mm}
	\subfloat[Vehicle speed and battery SoC comparison]{\includegraphics[width=\columnwidth]{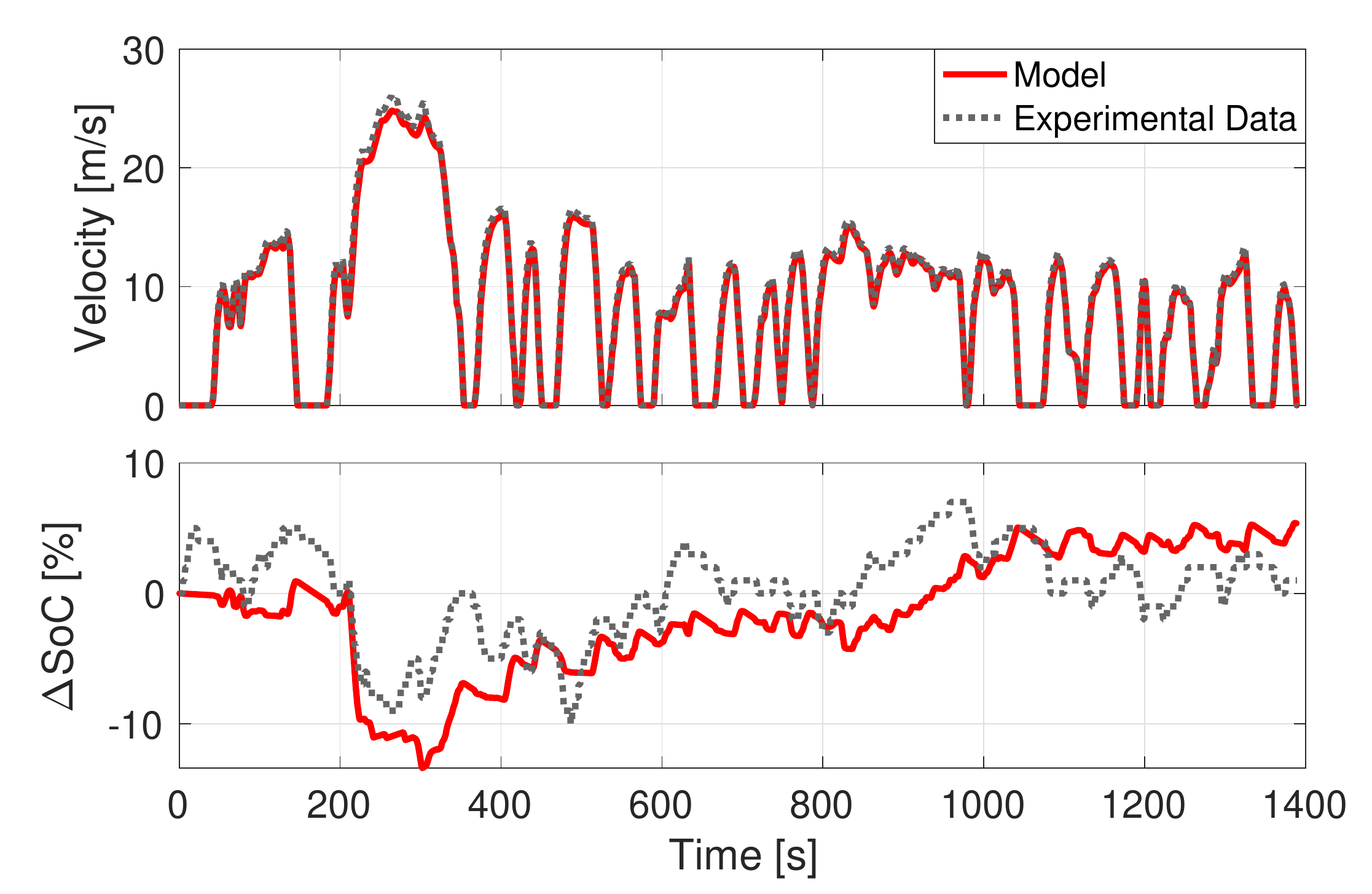}}
	\hfil
	\subfloat[Cumulative fuel consumption comparison]{\includegraphics[width=0.7\columnwidth]{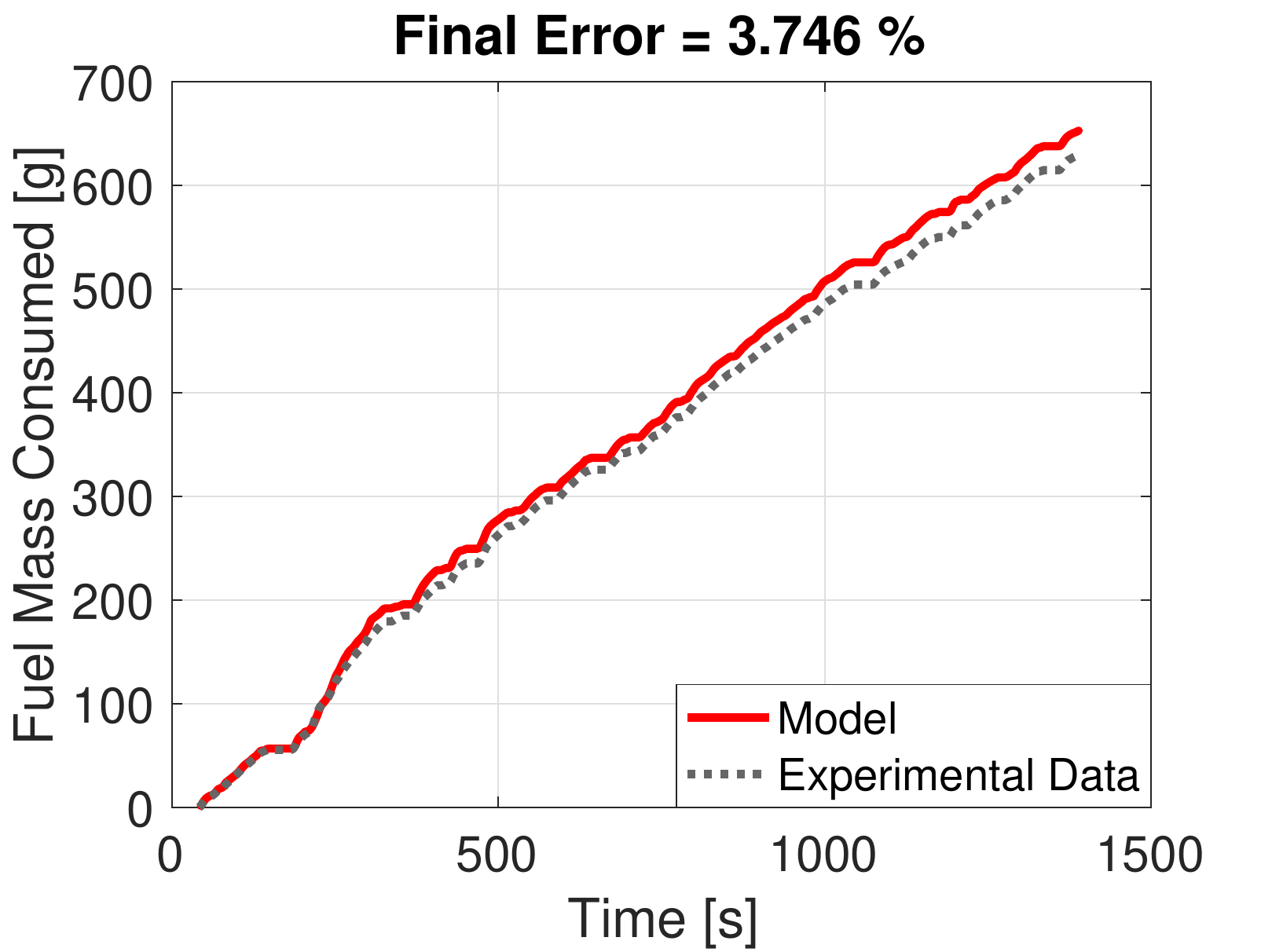}}
	\caption{Validation of forward vehicle model over FTP cycle.}
	\label{fig::model validation_FTP_results}
	\vspace{-1mm}
\end{figure}

The vehicle model was calibrated using experimental data collected from chassis dynamometer testing. The key variables used for evaluating the model are vehicle velocity, battery SoC, gear number, engine speed, desired engine and BSG torque profiles, and the fuel consumption. Fig. \ref{fig::model validation_FTP_results} shows sample results from model verification over the FTP regulatory drive cycle, where the vehicle velocity, battery SoC and fuel consumption are compared against experimental data.

The predicted vehicle velocity adequately matches the experimental data. Mismatches in the battery SoC profiles can be attributed to the simplicity of the battery model, in which electrical accessory loads are modeled using a constant current bias. The fuel consumption over the FTP cycle is well estimated by the model, with error on the final value less than $4\si{\%}$ relative to the actual engine. The error in the terminal SoC is \SI{3}{\%} on a \SI{48}{V}, $\SI{8}{Ah}$ battery. This is comparable to the SoC estimation error in vehicle. The calibration is considered satisfactory for energy and fuel consumption prediction over user-defined routes.

\section{Problem Formulation}
\label{sec::problem}

The objective of the nonlinear dynamic optimization problem, formulated in the spatial domain, is to minimize the fuel consumption of the vehicle over an entire itinerary. A key benefit of a spatial trajectory formulation is that it naturally lends itself to the incorporation of route-related information, such as posted speed limit signs, location of traffic lights and stop signs, whose positions along the route remain fixed.


Consider a dynamic control problem discretized in the spatial domain having the form:
\begin{equation}
x_{s+1} = f_s \left(x_s,u_s \right), \quad s = 1,\dots, N.
\end{equation}
where $s$ is the discrete position, $x_s \in \mathcal{X} \subset \mathbb{R}^p$ is the state, $u_s \in \mathcal{U} \subset \mathbb{R}^q$ is the input or control, and $f_s$ is a function that describes the state dynamics. In this work, the state variables chosen are the vehicle velocity and the battery SoC: $x_s = \left[v_s, \xi_s \right]^\mathsf{T}$. The engine torque and BSG torque are chosen as the control variables: $u_s = [T_{eng,s}, T_{bsg,s}]^\mathsf{T}$.

The control and the state are constrained, and the constraint function $h_s: \mathcal{X} \times \mathcal{U} \to \mathbb{R}^r$ is expressed as $h_s(x_s,u_s) \leq 0, \forall s = 1, \dots, N$, which includes the route speed limits, operating limits of physical actuators and subsystems, constraints for drive comfort and so on. An admissible control map at position $s$ is a map $\mu_s : \mathcal{X} \to \mathcal{U}$ such that $h(x,\mu_s(x)) \leq 0, \forall x \in \mathcal{X}$. The collection of admissible control maps is denoted by $\mathcal{M} := \left(\mu_{1}, \dots, \mu_{N} \right)$, which is referred to as the policy of the controller.

The controller aims at minimizing a cost, given by:
\begin{equation}
\label{eq::prb_cost_fn_gen}
\begin{aligned}
J(\mathcal{M}) = c_{N+1}(x_{N+1}) + \sum_{s = 1}^{N} c_s(x_s,u_s)
\end{aligned}
\end{equation}
where, $c_s : \mathcal{X} \times \mathcal{U} \to \mathbb{R}$ is the per stage cost function, defined in this work as a weighted average of the fuel consumption and travel time:
\begin{equation}
\label{eq::prb_cost_fn}
\begin{aligned}
c_s(x_s,u_s) = \left(\gamma \cdot \frac{\dot{m}_{f,s}(x_s,u_s)}{\dot{m}_f^{norm}} + (1-\gamma)\right) \cdot t_s \\
\end{aligned}
\end{equation}
where, the weight $\gamma \in (0,1)$ is a tunable penalty factor that can be used to trade-off between the amount of fuel consumed and time taken to complete the route; effectively it constitutes a driving aggressiveness parameter, $\dot{m}_{f,s}$ is the fuel consumption rate, $\dot{m}_f^{norm}$ is a cost normalizing weight and $t_s$ is the travel time per step.


\section{Full-Route Optimization using Dynamic Programming}
\label{sec::opt_soln_FR}

To perform the constrained optimization problem described in Section \ref{sec::problem}, a custom DP algorithm was developed and employed \cite{zhu2021gpu}. The DP formulation provides the closed-loop optimal policy, which makes it attractive for in-vehicle implementation. Specifically, for every feasible discretized state, there exists a control input that corresponds to the optimal strategy for the tail sub-problem. This property inherently adds robustness against unmodeled plant dynamics or other types of modeling errors. Further, DP can handle highly nonlinear cost and constraint functions while ensuring constraint satisfaction.

DP uses the \underline{B}ellman \underline{P}rinciple of \underline{O}ptimality (BPO) equation to break the optimization problem into smaller sub-problems, solving them via backward recursion. Mathematically, the proposition made in the DP algorithm is that for every initial state $x_1$, the optimal cost $J^*(\mathcal{M})$ of the problem defined in \eqref{eq::prb_cost_fn_gen} is equal to $J_1({x_1})$, given by the last step of the following algorithm, and solved using backward recursion from position $N$ to $1$:
\begin{equation*}
J_{N+1}(x_{N+1}) = c_{N+1}(x_{N+1}),
\end{equation*}
\begin{equation}
\label{eq::cost_fn_DP_FR}
\begin{aligned}
J_s(x_s) = \min_{\substack{\mu_s(x_s)}} \quad J_{s+1}\left(f_s(x_s,\mu_s(x_s)) \right) + c_s(x_s,\mu_s(x_s)),& \\
\quad \forall s = 1, \dots, N &
\end{aligned}
\end{equation}
Further, the policy $\mathcal{M}^* = \left(\mu_{1}^*,\dots, \mu_{N}^* \right)$ is optimal if for each $x_s $ and $s$, $\mu_s^*(x_s)$ minimizes the right side of \eqref{eq::cost_fn_DP_FR} \cite{bertsekas1995dynamic}. Here, $J_s(x_s)$ is interpreted as the optimal cost for the $(N+1-s)$-stage problem starting at state $x_s$ and position $s$, and ending at position $N+1$. For use in Section \ref{sec::opt_soln_RH}, the following assignment is made:
\begin{equation}
\label{eq::value_fn_defn}
\begin{aligned}
V_{N+1}(x_{N+1}) &= J_{N+1}(x_{N+1}), \\
V_s(x_s) &= J_s(x_s), \quad \forall  s = 1, \dots, N
\end{aligned}
\end{equation}
where $V_s$ is termed the value function, equal to the cost-to-go function at position $s$.

The constraints of the $N$-step optimization are defined as follows:
\begin{equation}
\label{eq::prb_constr_opt}
\begin{aligned}
v_{s} &\in \left[v_{s}^{min}, v_{s}^{max} \right], \quad \forall s = 2, \dots, N+1 \\
\xi_s &\in \left[\xi^{min}, \xi^{max} \right], \quad \forall s = 2, \dots, N+1 \\
v_1 &= v_1^{min}, \quad \xi_1 \in \left[\xi^{min}, \xi^{max} \right] \\
a_s &\in \left[a^{min}, a^{max} \right], \quad \forall s = 1, \dots, N \\
T_{eng,s} &\in \left[T_{eng,s}^{min}\left(v_s \right), T_{eng,s}^{max}\left(v_s \right) \right], \forall s = 1, \dots, N \\
T_{bsg,s} &\in \left[T_{bsg,s}^{min}\left(v_s \right), T_{bsg,s}^{max}\left(v_s \right) \right], \forall s = 1, \dots, N
\end{aligned}
\end{equation}
where, $v_s^{min}, v_s^{max}$ are the minimum and maximum route speed limits respectively; $\xi^{min}, \xi^{max}$ represent the static limits applied on battery SoC variation; $a^{min}, a^{max}$ are the limits imposed on the vehicle acceleration for comfort; $T_{eng,s}^{min}, T_{eng,s}^{max}$ are the state-dependent minimum and maximum torque limits of the engine respectively, and $T_{bsg,s}^{min}, T_{bsg,s}^{max}$ are the state-dependent minimum and maximum BSG torque limits respectively. To ensure SoC-neutrality over the global optimization, a terminal constraint is applied on the battery SoC: $\xi_1 = \xi_{N+1}$. Dynamical constraints are imposed by the vehicle model dynamics, described in \eqref{eq::longitudinal_dyn_model}-\eqref{eq::transmission_model}. Here, as the dynamic optimization problem is solved by the DP algorithm, the state dynamics introduced in \eqref{eq::longitudinal_dyn_model} and \eqref{eq::battery_model} are discretized and transformed to spatial domain:
\begin{equation}
\label{eq::state_equations_DP}
\begin{aligned}
v_{s+1}^2 &= v_s^2 + 2 \Delta d_s \cdot \biggl(\frac{F_{trc,s} - F_{road,s}(v_s)}{M} \biggr) \\
\xi_{s+1}  &= \xi_s - \frac{\Delta d_s}{\bar{v}_{s}}\cdot \frac{\bar{I}_{batt,s}}{C_{nom}}
\end{aligned}
\end{equation}
where $\Delta d_s$ is the distance over one step (i.e. $\Delta d_s = d_{s+1}-d_s$, where $d_s$ is the distance traveled along the route at position $s$) and $\bar{v}_s \left(= \frac{v_s + v_{s+1}}{2} \right)$ is the average velocity over one step.

In this formulation, the signal phase information of each traffic light is deterministically incorporated as part of the initialization process before the trip begins. Varying timing information (i.e. time in each phase) however, cannot be utilized in the full-route optimization routine. To address the unknown phase and timing of each traffic light in real-time, a receding horizon framework integrated with a pass-in-green model is proposed in this work, discussed in Section \ref{sec::PiGe_formulation}.

\begin{figure}[!t]
	\centering
	\vspace{-3mm}
	\subfloat[Pareto curve]{\includegraphics[width=0.8\columnwidth]{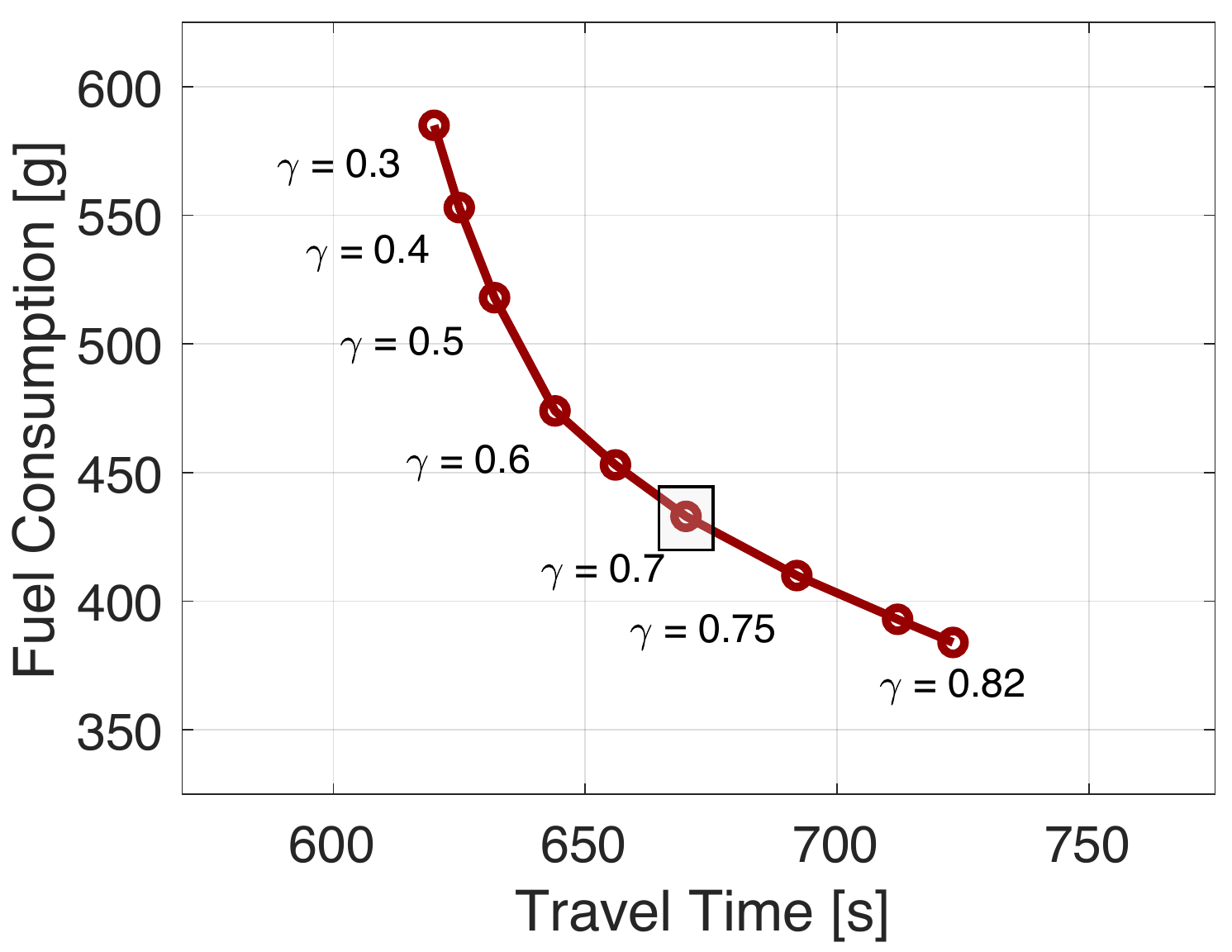}}
	\hfil
	\subfloat[Vehicle velocity and battery SoC profile, $\gamma = 0.7$]{\includegraphics[width=\columnwidth]{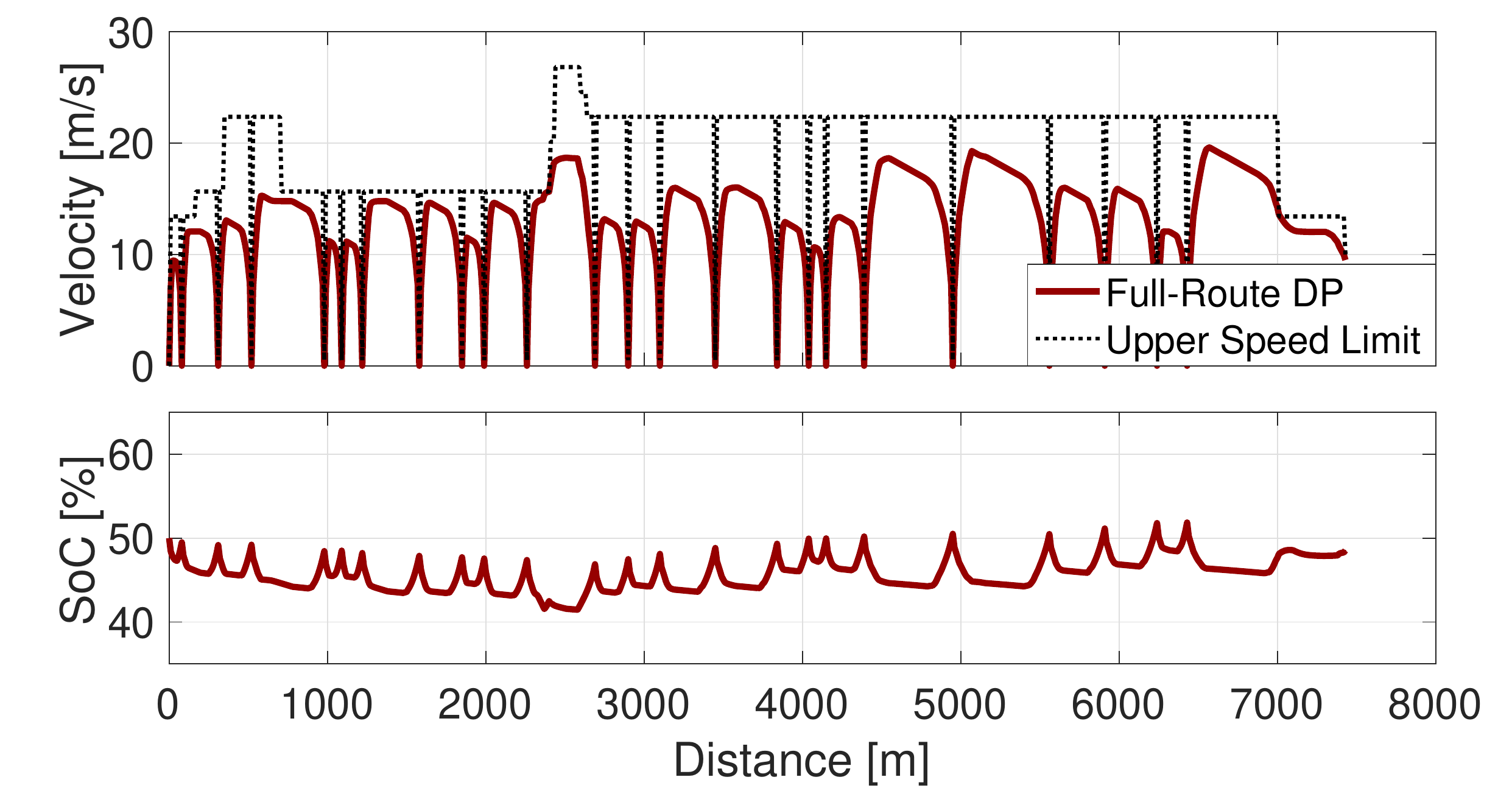}}
	\caption{Sample results from full-route DP optimization.}
	\label{fig::dpo_full_route_sample_results}
	\vspace{-1mm}
\end{figure}

Fig. \ref{fig::dpo_full_route_sample_results} shows sample results from a full-route DP optimization over an urban test route, in which all the traffic lights are assumed red (equivalent to stop signs). From Fig. \ref{fig::dpo_full_route_sample_results}(a), it is evident that the resulting solution of the multi-objective optimization problem \eqref{eq::prb_cost_fn} yields a $\gamma$-dependent Pareto front. Along the Pareto curve, lower values of $\gamma$ depict an increasingly aggressive driving style, while higher $\gamma$ values represent more conservative behavior with smoother accelerations and braking maneuvers. In Fig. \ref{fig::dpo_full_route_sample_results}(b), the optimal state trajectories reflecting a balanced scenario ($\gamma = 0.7$) are shown. The vehicle velocity profile is smooth and the torque split strategy determined by the optimization routine is charge-sustaining in nature.

\section{Model Predictive Control using Rollout Algorithm (Eco-Driving Algorithm)}
\label{sec::opt_soln_RH}

\begin{figure*}[!t]
	\centering
	\includegraphics[width=1.4\columnwidth]{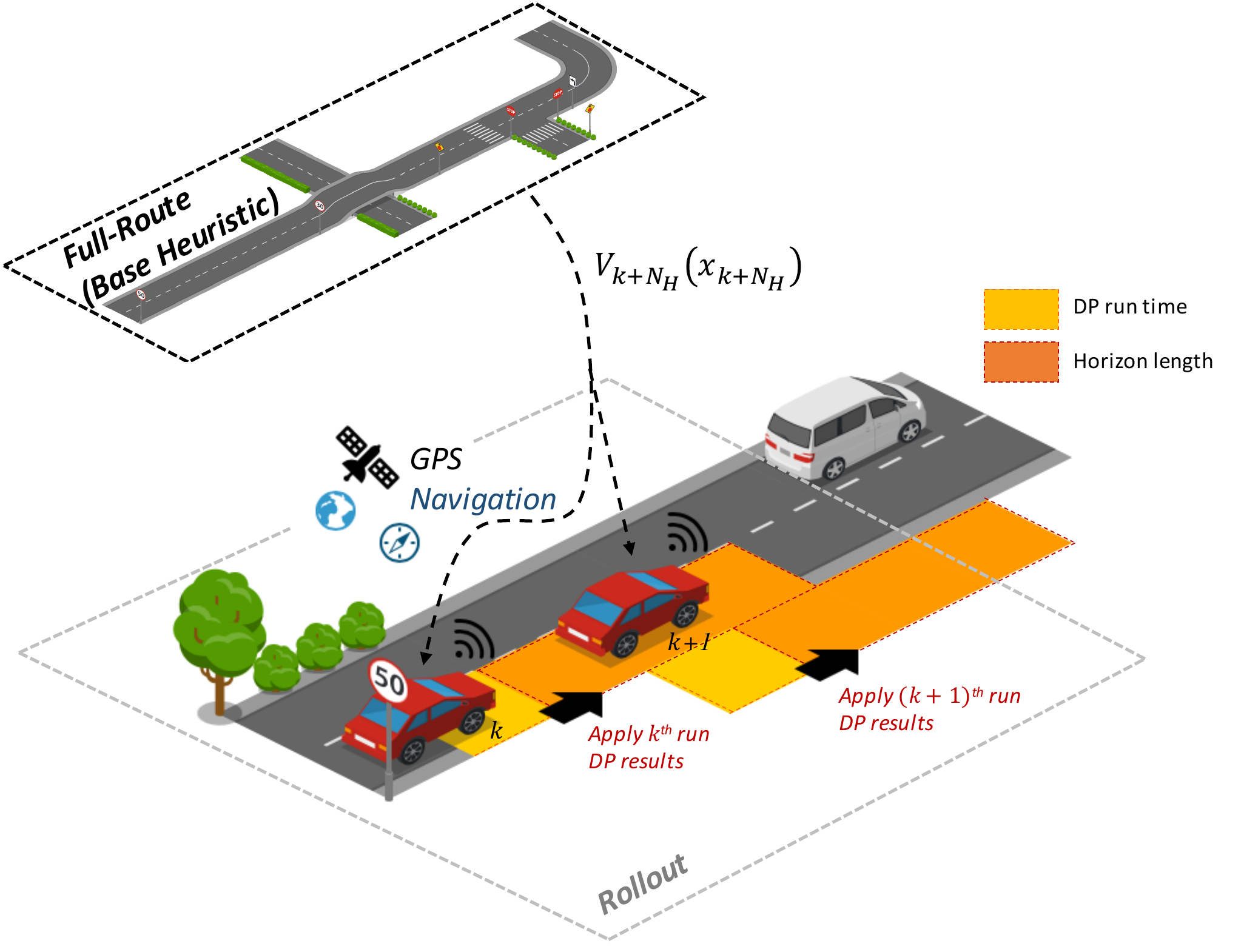}
	\caption{Illustration of rollout algorithm.}
	\label{fig::rollout_illustration}
\end{figure*}

If there is variability in route conditions and/or uncertainty in route information, the vehicle velocity and powertrain optimization will need to be re-run with the updated information to reflect these changes. For in-vehicle applications having limited computational resources, it becomes infeasible to periodically recompute the optimization for the remaining portion of the route online. This serves as the primary motivation to construct a receding horizon optimization problem by truncating the full-route horizon of $N$-steps to $N_H \ll N$ steps, formulated at a position $s$ as:
\begin{equation}
\label{eq::prb_cost_fn_RH}
\begin{aligned}
J_s^*(x_s) = \min_{\substack{\mathcal{M}^s}} \quad c_{s+N_H}(x_{s+N_H}) + \sum_{k = s}^{s+N_H-1} c_k(x_k,\mu_k(x_k)),& \\
\quad \forall s = 1, \dots, N-N_H+1, &
\end{aligned}
\end{equation}
\begin{equation*}
c_k(x_k,\mu_k(x_k)) =  \left(\gamma \cdot \frac{\dot{m}_{f,k}(x_k,\mu_k(x_k))}{\dot{m}_f^{norm}} + (1-\gamma)\right) \cdot t_k
\end{equation*}
where, $N_H$ is the number of steps in the receding horizon. This MPC or look-ahead optimization problem is subject to the same constraints introduced in \eqref{eq::prb_constr_opt}.


A key challenge in \eqref{eq::prb_cost_fn_RH} is the definition of an appropriate terminal cost and/or terminal state constraints that approximate the optimal solution provided by DP in a full-information scenario. This work introduces a terminal cost (or equivalently, the cost to complete the remaining route) approximation strategy based on the use of Approximate Dynamic Programming, specifically the rollout algorithm. Care has to be taken while imposing the terminal cost for the MPC -- a charge-sustaining constraint on the battery SoC over each horizon results in an overly conservative torque split strategy, while a greedy heuristic can lead to violation of SoC-neutrality over the entire trip. Some other methods adopted in the literature for constructing approximations of the value function include local linear approximation \cite{johannesson2008approximate,hellstrom2010design}, Monte Carlo-based \cite{bae2019real} and Q-learning-based approaches \cite{bertsekas1996neuro,melo2008analysis,zhu2020energy}.

Rollout algorithms form an important class of DP-based online suboptimal control techniques based on approximation in the value space. Here, the optimal cost-to-go function is replaced by an approximation, resulting in a suboptimal DP policy. Using the rollout algorithm framework, the following one-step look-ahead optimization problem is solved from position $k+N_H-1$ to $k, \quad  \forall k = 1, \dots, N-N_H+1$:
\begin{equation*}
\tilde{J}_{k+N_H}(x_{k+N_H}) = c_{k+N_H}(x_{k+N_H}),
\end{equation*}
\begin{equation}
\label{eq::cost_fn_DP_RH}
\begin{aligned}
\tilde{J}_s(x_s) = \min_{\substack{\hat{\mu}_s(x_s)}} \quad \tilde{J}_{s+1} \left(f_s(x_s,\hat{\mu}_s(x_s)) \right) + c_s(x_s,\hat{\mu}_s(x_s)),& \\
\quad \forall s = k, \dots, k+N_H-1 &
\end{aligned}
\end{equation}
where the approximation $\tilde{J}_{k+N_H}$ (and as a result $\tilde{J}_{s+1}$) is the cost-to-go of a known suboptimal policy, termed as the base policy or base heuristic, and $\mathcal{\hat{M}}^* := \left(\hat{\mu}_{k}^*, \dots, \hat{\mu}_{k+N_H-1}^* \right)$ is the rollout policy. One of the properties that make the rollout algorithm attractive for on-board optimization is the cost improvement property \cite{bertsekas2005rollout}, namely if the base heuristic produces a feasible solution, the rollout algorithm also produces a feasible solution whose cost is no worse than the cost corresponding to the base heuristic (proof in \cite{bertsekas1995dynamic}, adapted in Appendix \ref{app:cost_improvement_proof}).

For cost improvement to be valid, it is important that the base heuristic and the rollout policy are computed over the same constraint set. In the context of the real-time Eco-Driving problem considered in this work, cost improvement is of relevance as the rollout algorithm is inherently robust to parametric uncertainties and modeling errors experienced en route. Here, the base heuristic is chosen as the value function of the corresponding full-route DP:
\begin{equation*}
\begin{aligned}
\tilde{J}_{k+N_H}(x_{k+N_H}) = V_{k+N_H}(x_{k+N_H}), &\\
\quad \forall k = 1, \dots, N-N_H+1 &
\end{aligned}
\end{equation*}
where $V_k$ is the value function of the full-route DP solution at $k$, the global position along the route. The rationale behind this is explained using the BPO equation. Setting the terminal cost in \eqref{eq::cost_fn_DP_RH} to the value function of the full-route DP results in \eqref{eq::cost_fn_DP_FR}-\eqref{eq::value_fn_defn} for the $N_H$-step problem. Solving this system of equations thus yields the optimal cost for the look-ahead optimization problem. This claim is valid in the absence of traffic or other uncertainties en route i.e. as long as the per stage cost remains the same for both the $N_H$-step and $N$-step DP.

Fig. \ref{fig::rollout_illustration} is a visual representation of the online vehicle velocity and powertrain optimization framework developed in this work. The value function from the deterministic full-route DP optimization is computed at the beginning of the itinerary and applied as the terminal cost of each online receding horizon optimization. This base heuristic operates in conjunction with the policy improvement that at each step applies a correction to compensate for uncertainties en route and modeling errors.

The formulation of the Eco-Driving problem using the rollout algorithm is summarized as follows. At each global position $k$ along the route, the problem below is solved from local position $k+N_H-1$ to $k, \quad  \forall k = 1, \dots, N-N_H+1$:
\begin{equation*}
\tilde{J}_{k+N_H}(x_{k+N_H}) = V_{k+N_H}(x_{k+N_H}),
\end{equation*}
\begin{equation*}
\begin{aligned}
\tilde{J}_s(x_s) = \min_{\substack{\hat{\mu}_s(x_s)}} \quad \tilde{J}_{s+1} \left(f_s(x_s,\hat{\mu}_s(x_s)) \right) + c_s(x_s,\hat{\mu}_s(x_s)),& \\
\quad \forall s = k, \dots, k+N_H-1,&
\end{aligned}
\end{equation*}
\begin{equation*}
c_s(x_s,\hat{\mu}_s(x_s)) =  \left(\gamma \cdot \frac{\dot{m}_{f,s}(x_s,\hat{\mu}_s(x_s))}{\dot{m}_f^{norm}} + (1-\gamma)\right) \cdot t_s
\end{equation*}
subject to: $\quad \forall k = 1, \dots, N-N_H+1,$
\begin{equation}
\label{eq::prb_constr_opt_RH}
\begin{aligned}
v_{s} &\in \left[v_{s}^{min}, v_{s}^{max} \right], \quad \forall s = k+1, \dots, k+N_H \\
\xi_s &\in \left[\xi^{min}, \xi^{max} \right], \quad \forall s = k+1, \dots, k+N_H \\
v_1 &= v_1^{min}, \quad \xi_1 \in \left[\xi^{min}, \xi^{max} \right], \quad \xi_1 = \xi_{N+1} \\
a_s &\in \left[a^{min}, a^{max} \right], \quad \forall s = k+1, \dots, k+N_H \\
T_{eng,s} &\in \left[T_{eng,s}^{min}\left(v_s \right), T_{eng,s}^{max}\left(v_s \right) \right], \forall s = k, \dots, k+N_H-1 \\
T_{bsg,s} &\in \left[T_{bsg,s}^{min}\left(v_s \right), T_{bsg,s}^{max}\left(v_s \right) \right], \forall s = k, \dots, k+N_H-1
\end{aligned}
\end{equation}

\subsection{Incorporation of SPaT Information (Eco-Driving Algorithm with Eco-AND)}
\label{sec::PiGe_formulation}

Incorporating SPaT information in the aforementioned Eco-Driving framework can be addressed by adding time as a state variable in the optimization problem formulation \cite{sun2020optimal}. However, this is accompanied by an exponential increase in the computational effort required to compute the optimal solution \cite{larson1978principles}.

In this work, a computationally economical alternative is proposed to perform Eco-AND at signalized intersections by utilizing V2I technologies. Specifically, a rule-based approach is developed to increase the likelihood of passing-in-green by using SPaT information broadcast to the vehicle within communication range to determine kinematically feasible vehicle velocity constraints that are then imposed to the MPC routine.

The feasible velocity constraints are applied as offsets to the route speed limits, and are determined by first examining the possibility to execute a pass-in-green maneuver while respecting vehicle dynamics constraints and route speed limits. If deemed infeasible, the traffic light is treated as a stop sign and the vehicle decelerates to a stop according to the fuel-optimal profile computed by the rollout algorithm. If a pass-in-green maneuver is deemed feasible, the Eco-AND algorithm uses the current vehicle velocity, distance to the upcoming intersection, and the SPaT information to calculate the offsets $v_{off,s}^{min}$ or $v_{off,s}^{max}$ that are applied to the minimum and/or maximum route speed limit respectively. The effective velocity constraints fed to the MPC are summarized below:
\begin{equation}
\label{eq::EcoAND_speed_lim_constr}
\begin{aligned}
v_{s} \in \left[v_{s}^{min} + v_{off,s}^{min}, v_{s}^{max} - v_{off,s}^{max} \right], &\\
\quad \forall s = k+1, \dots, k+N_H &
\end{aligned}
\end{equation}
However, changing the vehicle speed constraints abruptly from the route speed limits to those prescribed by the Eco-AND algorithm will likely cause an infeasibility in the optimization routine. To ensure a feasible solution, the constraints fed to the rollout algorithm are suitably shaped by recursively applying: $v_{s+1}^2 = v_s^2 - 2a^{min} \Delta d_s$, from the current speed till the calculated (modified) speed limit is reached. Constraint shaping in this manner increases the likelihood with which the vehicle crosses the intersection in the green window while ensuring kinematic feasibility for use in the optimization routine.

\section{Verification Framework}

\begin{figure}[!t]
	\centering
	\includegraphics[width=\columnwidth]{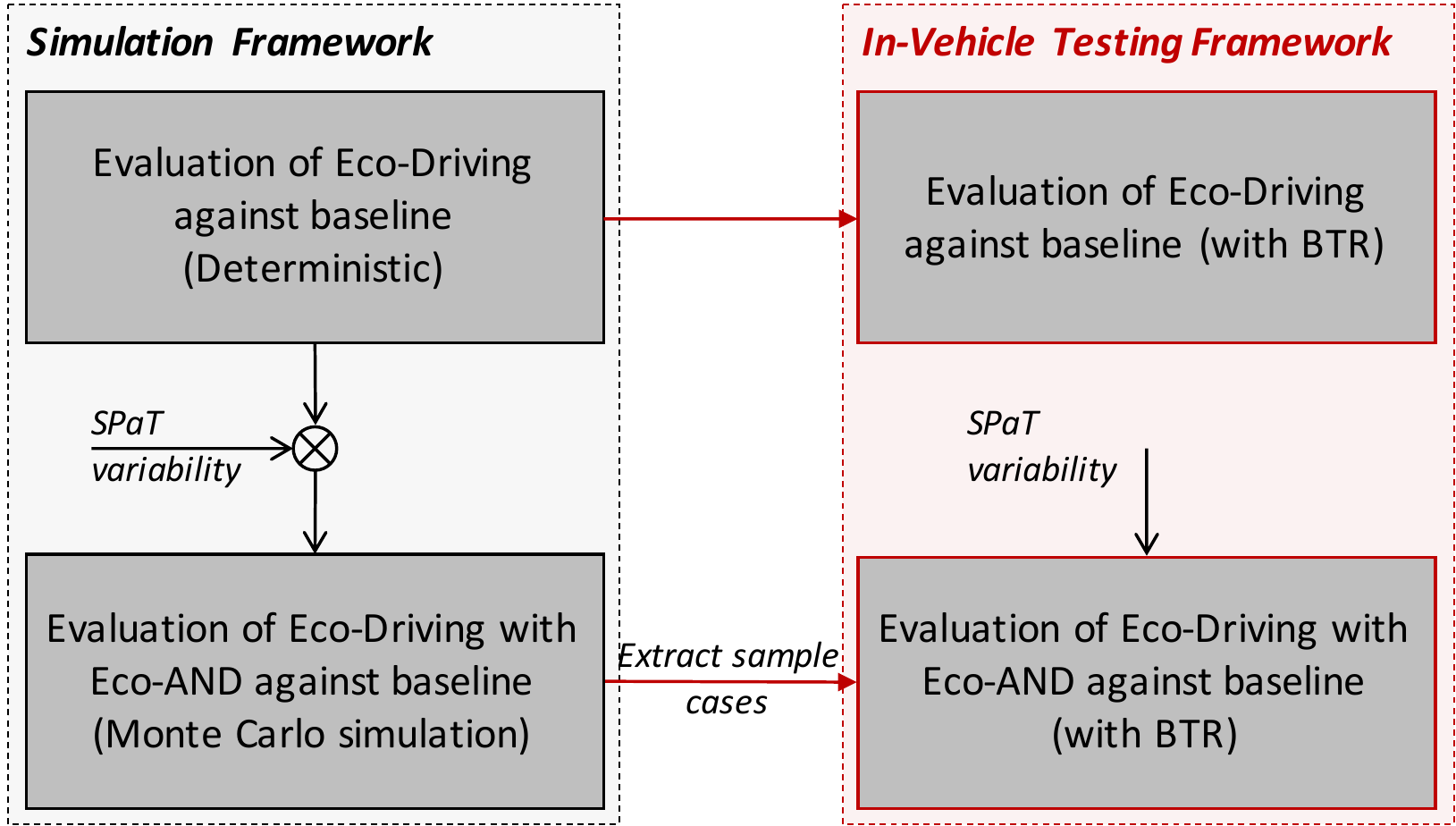}
	\caption{Process for verification of proposed optimization strategies.}
	\label{fig::verification_frmwrk_flow}
\end{figure}

The process for verification of the proposed vehicle velocity and powertrain optimization strategy is summarized in Fig. \ref{fig::verification_frmwrk_flow}. The virtual Evaluation of the Eco-Driving case involves the generation of multiple simulation scenarios, where the parameter $\gamma$ in the optimization is varied to determine and quantify the Pareto-optimal fronts among the objectives (total fuel consumption and trip travel time). This evaluation is initially performed by assuming that all the traffic lights along the route are stop signs. The optimizer is benchmarked against a realistic baseline, which is assumed as the same demonstration vehicle without DSF. Note that for the baseline case, the $\SI{48}{V}$ mHEV system is enabled. Further, no longitudinal automation is assumed, and for this reason all the simulations were conducted by including a validated \underline{E}nhanced \underline{D}river \underline{M}odel (EDM), a modified car-following model that mimics the behavior of a human driver in presence of speed limits and signalized intersections \cite{gupta2019enhanced}.

Following this initial virtual verification, experimental testing was conducted on a closed test track at the \underline{T}ransportation \underline{R}esearch \underline{C}enter (TRC) Inc. in East Liberty, OH, where the Eco-Driving algorithm was demonstrated through real-time implementation in the demonstration vehicle. To obtain a realistic baseline for benchmarking the results, the vehicle was fitted with a \underline{B}rake and \underline{T}hrottle \underline{R}obot (BTR) \cite{coovert2009design}, programmed to follow pre-established velocity profiles generated in simulation with the EDM. This deterministic scenario was used to validate the simulation tools on specific scenarios, offering an initial estimate of the fuel-saving potentials of the developed Eco-Driving algorithm.

Next, a more comprehensive virtual verification (Evaluation of Eco-Driving algorithm with Eco-AND, in Fig. \ref{fig::verification_frmwrk_flow}) was performed using a Monte Carlo simulation framework, in which the driver aggressiveness and the SPaT information are treated as random variables. From the results of this simulation, in-vehicle verification at TRC was conducted by extracting and testing sample cases from corresponding scenarios selected from the Monte Carlo simulations. Here, SPaT variability, including communication latencies, was realistically emulated during testing. Benchmarking was performed using a BTR that follows the velocity profiles generated in simulation by the EDM, where the decision to approach a signalized intersection is based on the concept of \underline{L}ine-\underline{o}f-\underline{S}ight (LoS) \cite{rajakumar2020benchmarking}. Choosing a simulation model to generate the reference velocity profile for vehicle testing guarantees repeatability of the experiment and limits the uncertainties induced by the testing environment.

\section{Virtual Evaluation}

\subsection{Test Route}

\begin{figure}[!t]
	\centering
	\includegraphics[width=\columnwidth]{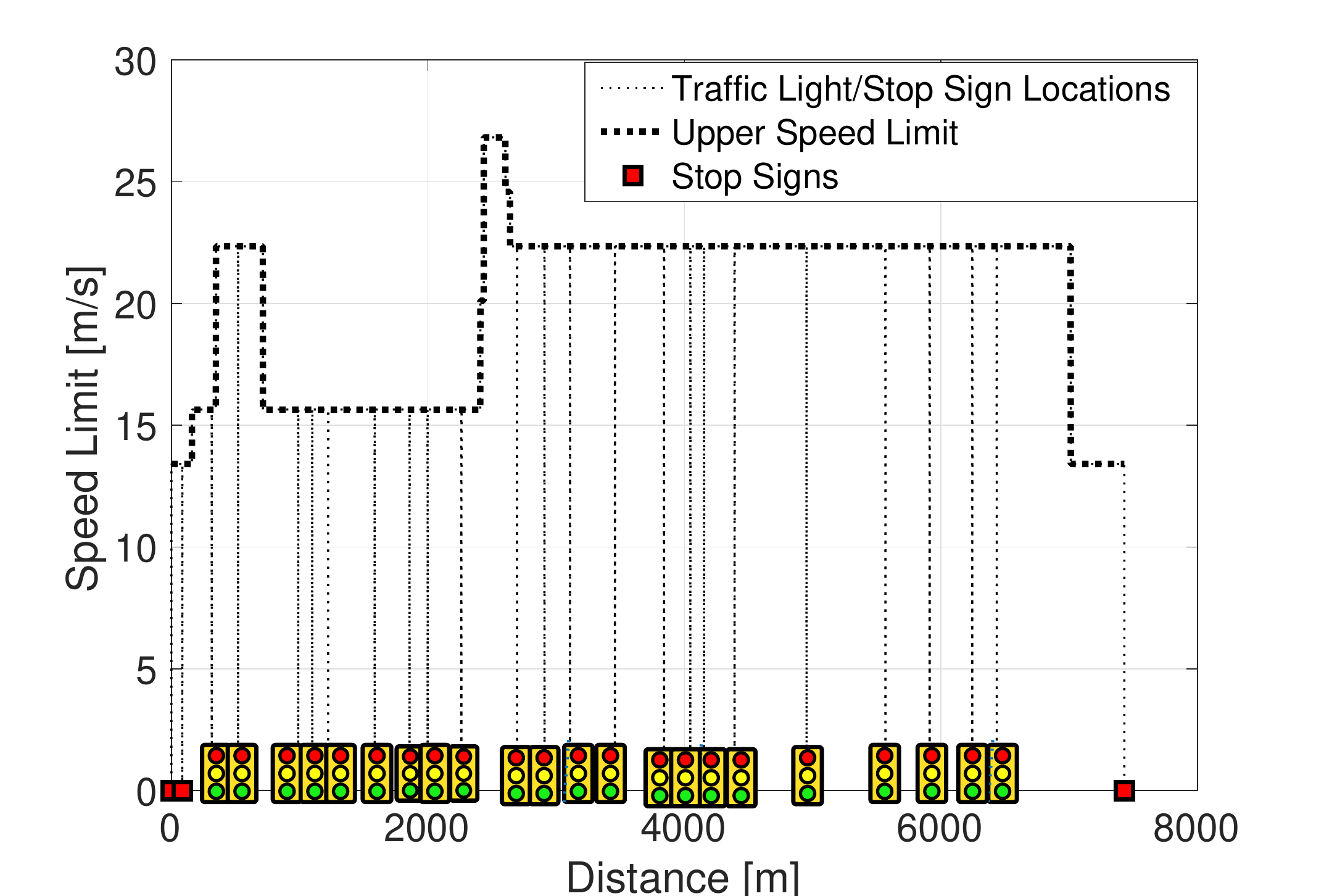}
	\caption{Speed limits and locations of traffic lights and stop signs along the selected route.}
	\label{fig::R_15_features}
\end{figure}

Fig. \ref{fig::R_15_features} shows the features of the representative route over which the optimization routine is tested. This urban route set in Columbus, OH is $\SI{7.4}{km}$ in length and comprises $22$ traffic lights and $3$ stop signs.

\subsection{Evaluation of Eco-Driving Algorithm}

\begin{figure}[!t]
	\centering
	\includegraphics[width=0.85\columnwidth]{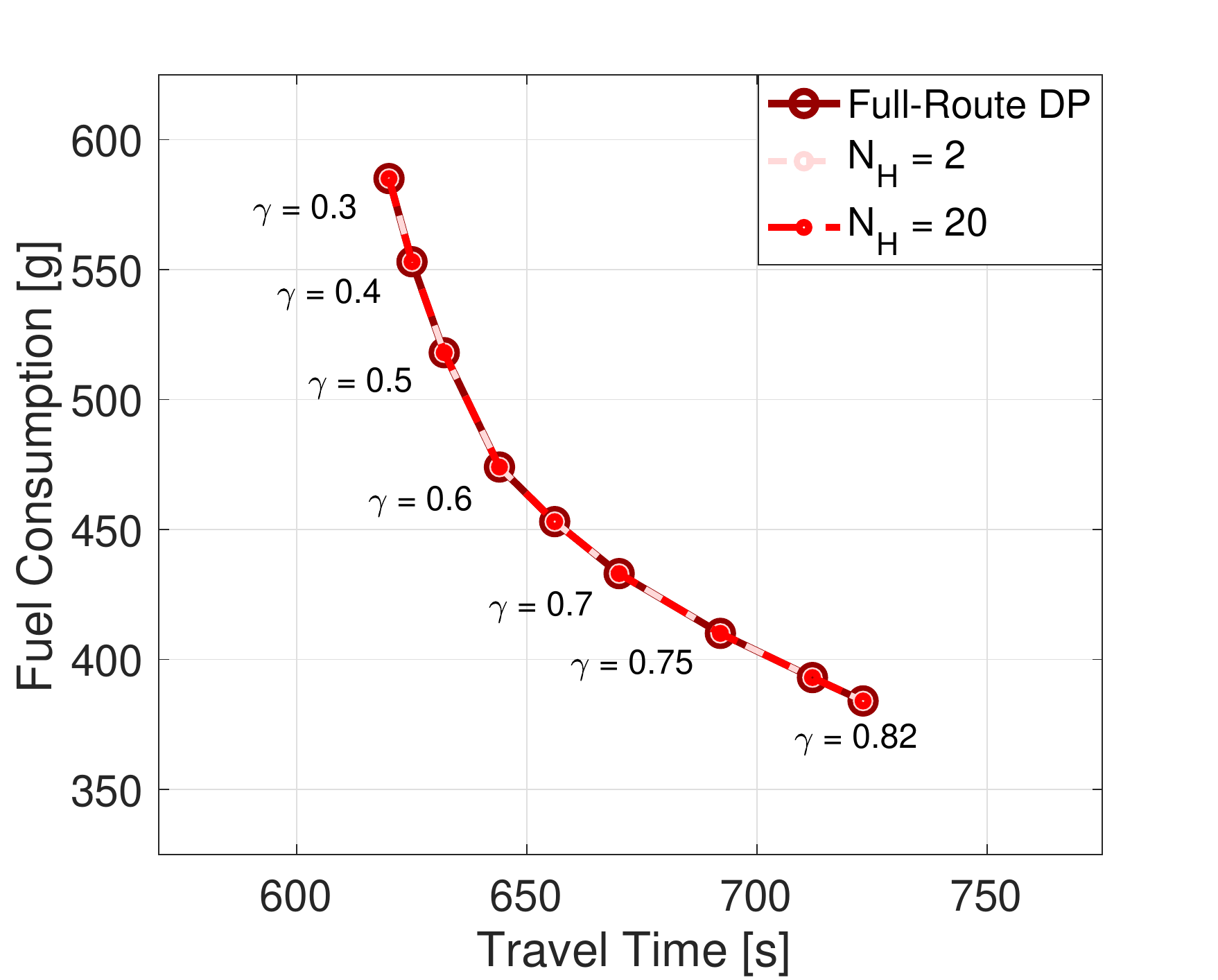}
	\caption{Sample results from the Eco-Driving algorithm: Comparison of Pareto curve with the full-route DP.}
	\label{fig::dpo_rollout_sample_results}
\end{figure}

Fig. \ref{fig::dpo_rollout_sample_results} shows sample results from the Eco-Driving algorithm over the urban test route with no traffic, in which all the traffic lights are assumed to be stop signs. These results are in accordance with the BPO equation $\forall N_H$ (and the claim made in Section \ref{sec::opt_soln_RH}), as evident from the comparison with the Pareto curve from the full-route DP optimization over the same route. This result is obtained by imposing identical constraints on the full-route DP optimization and the MPC (Eco-Driving algorithm), and the value function from the full-route DP solution is applied as the terminal cost of each $N_H$-horizon problem. The step sizes of the discretized distance, state and control grids used for simulation are provided in Table \ref{tab::DP_grid_discretization} (selection was made after repeated simulations and based on the desired accuracy and associated computational requirements for this specific system and application).


To highlight the role of cost improvement in the rollout algorithm, a realistic parametric variability is introduced in the optimization problem. Specifically, let the total vehicle mass (or alternately the payload) that is considered for performing the full-route DP optimization and computing its value function be different from the actual payload carried en route. This can be interpreted as a scenario in which the number of passengers and their mass cannot be accurately determined before the trip begins. To demonstrate cost improvement using simulations in this case, the value function used in the Eco-Driving algorithm is computed using the pre-trip (or original) vehicle mass while the stage cost is computed using the en route (or perturbed) mass of the vehicle (assuming that the total vehicle mass can be estimated online, as seen in \cite{fathy2008online,vahidi2005recursive}). The results of this MPC are then compared with that obtained by full-route DP optimization using the original mass of the vehicle. Further, it is also compared with the global optimum i.e. a full-route DP solution using the perturbed mass of the vehicle (that in a realistic sense cannot be determined a priori). Note that the choice of grid discretization used for this evaluation is provided in Table \ref{tab::DP_grid_discretization}.

%

To quantitatively evaluate the performance of the different optimization routines, the cumulative costs that they accrue are compared using the following definition:
\begin{equation}
\label{eq::cum_cost_metric_eval}
\begin{aligned}
J^*(x_1) &= \sum_{k = 1}^{N} \left(\gamma \cdot \frac{\dot{m}_{f,k}^*}{\dot{m}_f^{norm}} + (1-\gamma)\right) \cdot t_k^* \\
&= \gamma \cdot \frac{m_{f,N}^*}{\dot{m}_f^{norm}} + (1-\gamma) \cdot \tau_N^*
\end{aligned}
\end{equation}
where $m_{f,N}^*,\tau_N^*$ are respectively the resulting cumulative fuel consumption and total travel time from applying the optimal control policy. For fair evaluation, the cumulative costs are compared while ensuring battery SoC neutrality over their respective trips.

\begin{table}[!t]
	\centering
	\begin{tabular}{cc}
		\hline
		Discretized Variable & Step Size \\ \hline
		Distance  &  $\SI{10}{m}$ \\
		Velocity & $\SI{1.36}{m/s}$ \\
		Battery SoC & $\SI{2}{\%}$ \\
		Engine torque & $\SI{13.2}{Nm}$ \\
		BSG torque & $\SI{4.2}{Nm}$
	\end{tabular}
	\caption{Choice of grid discretization in DP.}
	\label{tab::DP_grid_discretization}
\end{table}

\begin{figure*}[!t]
	\centering
	\vspace{-3mm}
	\subfloat[Cost improvement in rollout algorithm, $N_H = 20$]{\includegraphics[width=0.8\columnwidth]{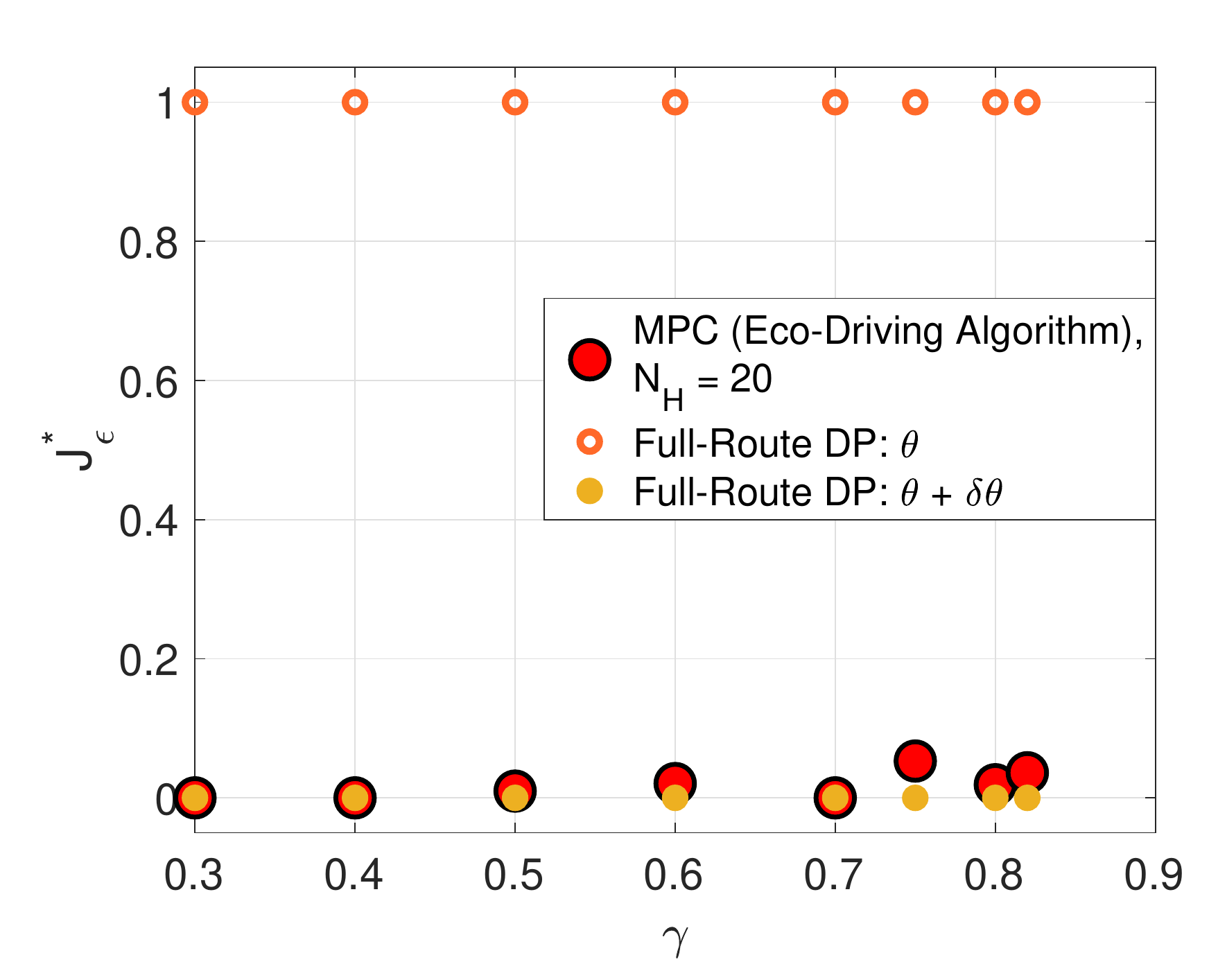}} 
	\subfloat[Impact of $N_H$ on cost improvement]{\includegraphics[width=1.2\columnwidth]{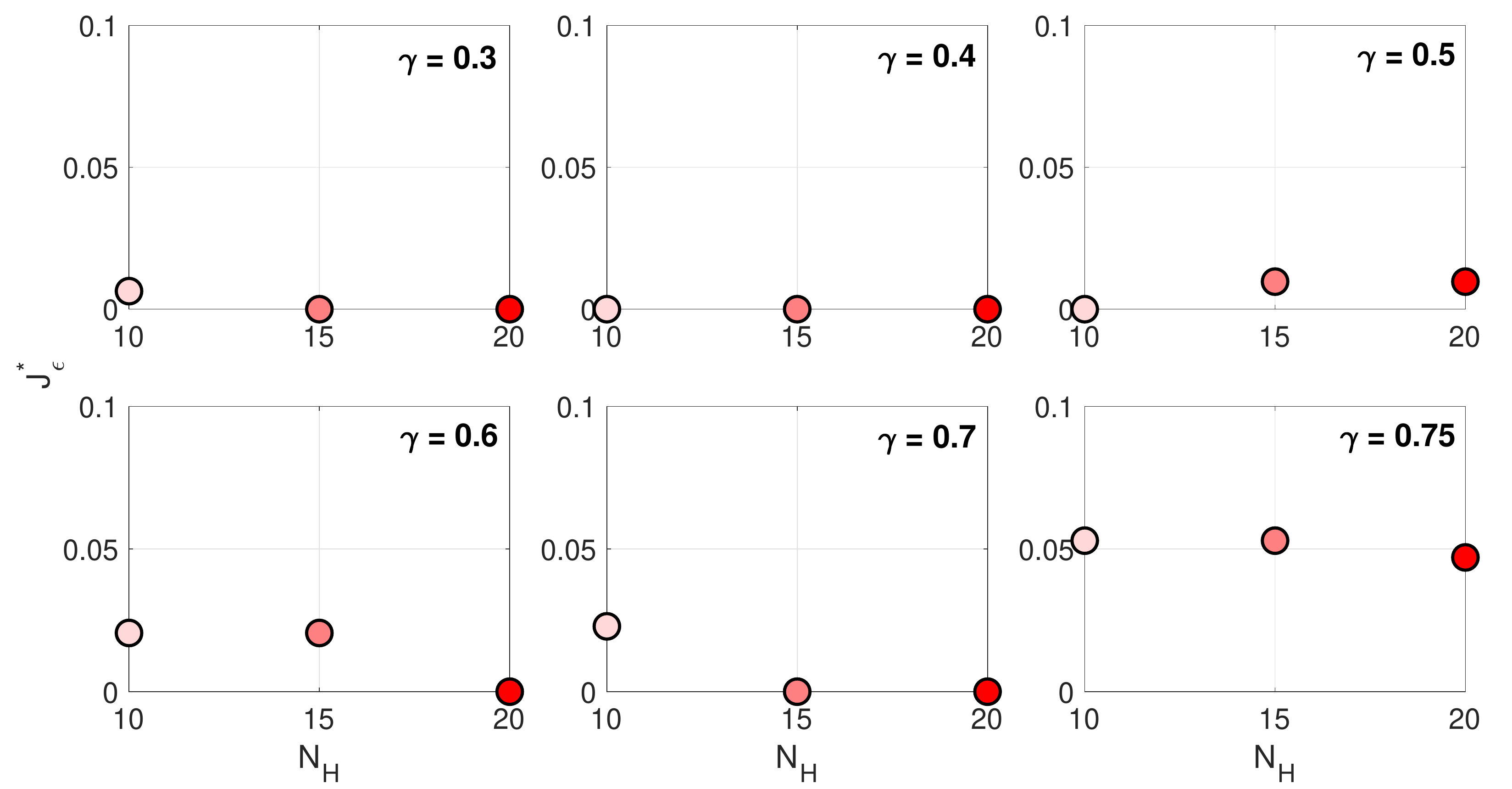}}
	\caption{Sample results on cost improvement from the Eco-Driving algorithm.}
	\label{fig::dpo_rollout_cost_imprvmnt}
	\vspace{-1mm}
\end{figure*}

Fig. \ref{fig::dpo_rollout_cost_imprvmnt}(a) highlights the cost improvement properties of the rollout algorithm framework proposed by introducing a significant parametric variation of $\SI{20}{\%}$, in which an error metric $J_\epsilon^*$ is compared for the different optimization routines over a range of $\gamma$. $J_\epsilon^*$ represents a normalized cumulative cost error term defined as follows:
\begin{equation}
\label{eq::cum_cost_metric_eval_cost_imprvmnt}
\begin{aligned}
J_\epsilon^*(x_1) = \abs*{\frac{J^*(x_1) - J_{\theta+\delta \theta}^*(x_1)}{J_{\theta}^*(x_1) - J_{\theta+\delta \theta}^*(x_1)}}
\end{aligned}
\end{equation}
where $J_{\theta}^*, J_{\theta+\delta \theta}^*$ are the cumulative costs accrued by the full-route DP optimization run with the original, and the perturbed (or optimum) vehicle mass respectively. Clearly, for $J^*(x_1) = J_{\theta+\delta \theta}^*(x_1), J_\epsilon^*(x_1) = 0$ and for $J^*(x_1) = J_{\theta}^*(x_1), J_\epsilon^*(x_1) = 1$, forming the bounds of the error term in Fig. \ref{fig::dpo_rollout_cost_imprvmnt}(a). The results from the numerical simulations are in excellent agreement with the cost improvement property of the rollout algorithm, with over $\SI{95}{\%}$ reduction in the error metric (compared to the full-route DP with the pre-trip mass).

Fig. \ref{fig::dpo_rollout_cost_imprvmnt}(b) shows the impact of the horizon length $N_H$ on the cost improvement results, in which the same error metric $J_{\epsilon}^*$ is used for comparison. It is seen that the error (w.r.to the perturbed full-route DP) progressively decreases as $N_H$ increases. This is expected as the role of the ``incorrect" terminal cost in determining the cost-to-go \eqref{eq::cost_fn_DP_RH} effectively decreases as the length of the receding horizon increases. The small discrepancy in this trend for $\gamma = 0.5$ can be attributed to minor numerical errors arising due to the particular choice of the horizon length, and discretization of the distance, state and control input grids.

\subsection{Evaluation of Eco-Driving Algorithm with Eco-AND}

The Eco-Driving algorithm with Eco-AND is evaluated by performing large-scale Monte Carlo simulations over real-world routes in which each traffic light has time-varying SPaT information.

\begin{figure*}[!t]
	\centering
	\includegraphics[width=2\columnwidth]{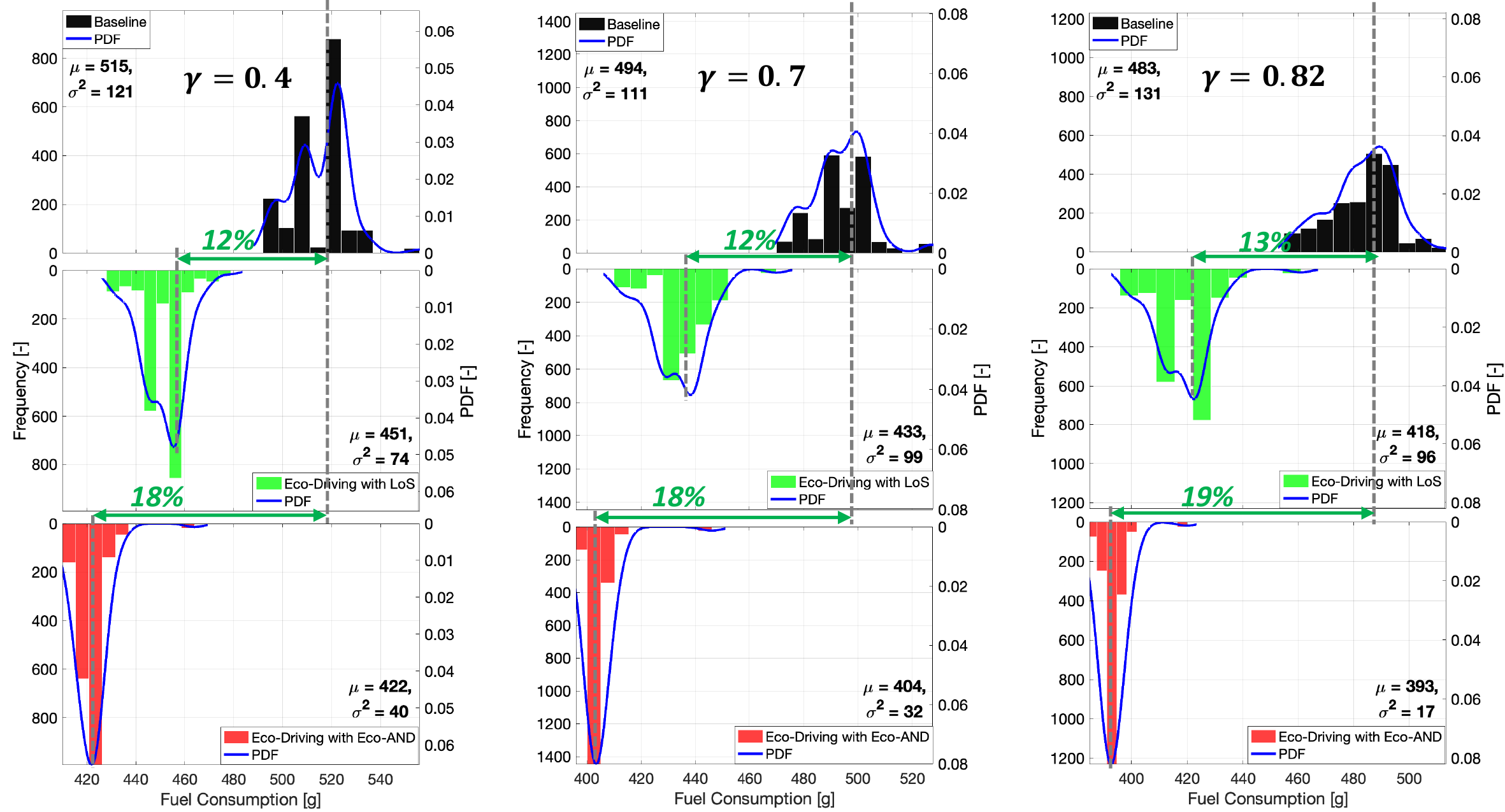}
	\caption{Comparison of fuel consumption between baseline, Eco-Driving with LoS and Eco-Driving with Eco-AND cases (urban test route, $\gamma = \{0.4,0.7,0.82 \}$, all results at same travel time).}
	\label{fig::MC_sim_results_R15}
\end{figure*}

\subsubsection{Baseline Case}

The baseline considered is the Enhanced Driver Model or EDM, a deterministic reference velocity predictor that uses route characteristics to generate velocity profiles representing different levels of driver aggressiveness \cite{gupta2019enhanced}. The EDM predicts the response of a human driver when operating a vehicle in presence of traffic, stop signs and traffic lights by using a realistic LoS based heuristic strategy \cite{rajakumar2020benchmarking}. For simulations over urban and mixed routes, a LoS of $\SI{100}{m}$ is considered to be reasonable. The velocity reference from the EDM is fed to a tracking controller that generates the necessary inputs to the validated vehicle model described in Section \ref{sec::model}.

An additional case is now constructed to quantify the fuel saving potentials of Eco-AND. The LoS approach (as introduced in \cite{rajakumar2020benchmarking}) is adapted and used in conjunction with the rollout algorithm to interact with traffic signals. Here, the signal phase information is available to the driver only within the LoS, and at any point beyond it, the traffic light is assumed to be a stop sign. It should be noted that unlike Eco-AND, the Eco-Driving with LoS does not receive any timing information from the traffic light.

\subsubsection{Variability in SPaT Information}

To recreate the degree of variability typically associated to SPaT at subsequent signalized intersections, a base SPaT sequence is extracted from the traffic simulation program SUMO (\underline{S}imulation of \underline{U}rban \underline{Mo}bility) \cite{behrisch2011sumo}. SUMO contains models of traffic lights in which the SPaT sequences are assigned based on the location, speed limits of the intersecting roads, presence of adjacent traffic lights and other factors \cite{krajzewicz2002sumo}. Once the desired route is created, the SPaT information and initial signal phase corresponding to a given lane at each signalized intersection along the route is obtained. The extracted SPaT sequences are then validated to be in accordance with the regulations outlined in the City of Columbus Traffic Signal Design Manual \cite{citycbus2021tsdm}. Starting from this base SPaT, all the traffic lights are offset by the same uniformly random value such that the phase difference between the traffic lights remains constant i.e. the signals are synchronized. Alternately, this can be interpreted as considering the departure time as a uniform random variable: $t_{dep} \sim \mathcal{U}\left(0,t_{cyc} \right)$, where $t_{dep}$ is the departure time and $t_{cyc}$ is the traffic cycle time, chosen in this study as $\SI{90}{s}$ (fixed).

\subsubsection{Monte Carlo Simulations}

Monte Carlo simulations are then performed over the test route by considering the SPaT information as a random variable. Here, $2000$ different scenarios are generated by randomly changing the departure time (as described above). The different cases being compared in this simulation study are: Baseline (EDM with LoS of $\SI{100}{m}$), Eco-Driving with LoS (no timing information, current phase of each traffic signal is known within a LoS of $\SI{100}{m}$, $N_H = 20$), and Eco-Driving with Eco-AND ($N_H = 20$). For both the Eco-Driving cases, three calibrations are considered: $\gamma = \{0.4,0.7,0.82 \}$, to represent an aggressive, normal and relaxed driving behavior respectively. Note that the choice of grid discretization used for this evaluation is shown in Table \ref{tab::DP_grid_discretization}. For fair comparison, the baseline EDM for each simulation is calibrated to represent a driver aggressiveness comparable to that value of $\gamma$. In total, $6000$ simulations are executed for each case.

Fig. \ref{fig::MC_sim_results_R15} shows the distributions of the fuel consumption for the baseline (\tikz \draw[white,fill=black] (0,0) circle (.5ex);), Eco-Driving with LoS (\tikz \draw[white,fill=green] (0,0) circle (.5ex);) and Eco-Driving with Eco-AND (\tikz \draw[white,fill=red] (0,0) circle (.5ex);) cases, corresponding to $\gamma = \{0.4,0.7,0.82 \}$. It is to be noted that the travel times obtained for each $\gamma$ remains the same across the three cases being evaluated. A non-parametric \underline{p}robability \underline{d}ensity \underline{f}unction (PDF) known as the \underline{K}ernel \underline{D}ensity \underline{E}stimator (KDE) is used to obtain the fitted distribution for each of the cases shown.

Over the mixed test route, it is seen that the Eco-Driving algorithm with Eco-AND reduces the fuel consumption of the baseline $\SI{48}{V}$ hybrid by $\SI{18}{\%}$, $\SI{18}{\%}$ and $\SI{19}{\%}$ corresponding to $\gamma = \{0.4,0.7,0.82 \}$ respectively. Further, there is a significant reduction in the standard deviation of the fuel consumption distribution compared to the baseline. This is a notable result as this means that the Eco-Driving algorithm equipped with Eco-AND can more consistently achieve a fuel consumption close to the mean of the distribution i.e. the spread of the fuel consumption due to variability in the driver aggressiveness and SPaT information is significantly reduced.

To quantify the benefits obtained from Eco-AND specifically, the results from Eco-Driving with LoS are compared against that with Eco-AND. For the urban route with $22$ traffic lights, the Eco-AND provides an incremental $\SI{6}{\%}$ fuel saving over the LoS implementation, with similar mean travel time. Simulations were also performed over other urban and mixed routes, the Eco-AND results in $2-10\si{\%}$ additional fuel savings. The reason for the increased benefits from the pass-in-green algorithm over urban routes can be attributed to the higher density of traffic lights.


\section{In-vehicle Implementation and Results}

Experimental tests were conducted at TRC to verify the implementation of the optimizer, and compare the results obtained by the demonstration vehicle against the simulation models and a realistic baseline. The tests conducted allow one to evaluate the real-world benefits of fuel consumption and travel time of the optimizer over reconstructed routes.

The MPC framework is implemented on the demonstration vehicle, a $2016$ VW Passat $\SI{1.8}{L}$ with a 6-speed automatic transmission and turbocharged gasoline engine, which was retrofitted as a mHEV by installing a BSG and a $\SI{48}{V}$ battery pack. The demonstration vehicle was also equipped with CAV technologies and DSF (shown in Fig. \ref{fig::NEXTCAR_TRC_images}(a)). For all baseline testing the $\SI{48}{V}$ mild-hybrid system is active and DSF is disabled, while for all optimizer testing both the mild-hybrid system and DSF are enabled. The on-board CAV technologies include an advanced GPS module for enhanced navigation, a \underline{D}edicated \underline{S}hort \underline{R}ange \underline{C}ommunication (DSRC) module that enables V2X communication, camera and radar modules to support \underline{A}daptive \underline{C}ruise \underline{C}ontrol (ACC) functionality. These CAV technologies enable the ability of full longitudinal control of the vehicle to achieve SAE Level $1+$ functionality (in accordance with SAE J3016).


The real-time Eco-Driving algorithm (with Eco-AND), integrated with V2X communication and ACC, have been implemented using rapid prototyping hardware (dSPACE MicroAutoBox II, or MABx in short) in the test vehicle. Online calibration of control parameters and data logging are performed using dSPACE ControlDesk software via the Host Ethernet interface. A significant result to be noted here is that the implemented optimizer can compute the solution to a $20$-step receding horizon optimization problem via the $2$-state, $2$-input DP in $\approx \SI{200}{ms}$ on the MABx.

\subsection{Experimental Test Setup}

\begin{figure}[!t]
	\centering
	\vspace{-3mm}
	\subfloat[NEXTCAR test vehicle]{\includegraphics[width=0.85\columnwidth]{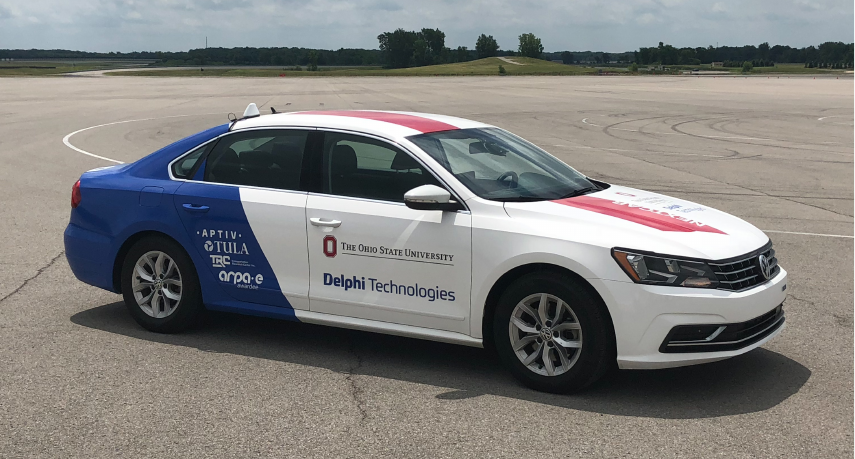}}
	\hfil
	\subfloat[Aerial view of High-Speed Test Track at TRC]{\includegraphics[width=\columnwidth]{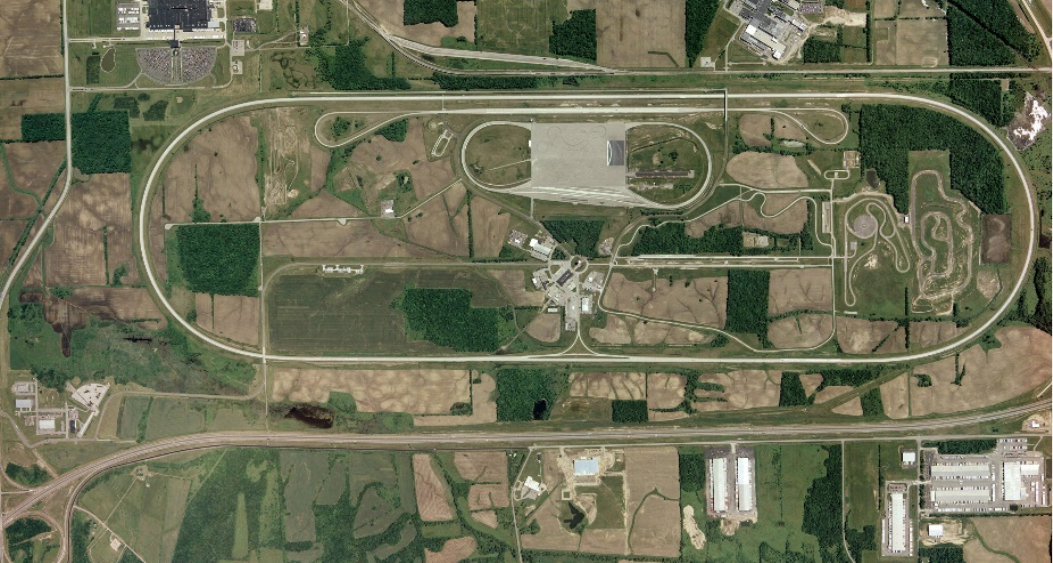}}
	\caption{Images of demonstration vehicle and experimental testing location.}
	\label{fig::NEXTCAR_TRC_images}
	\vspace{-1mm}
\end{figure}

All vehicle tests were performed in a single lane with $\SI{0}{\%}$ grade on the $\SI{7.5}{mi}$ High-Speed Test Track located at TRC (refer Fig. \ref{fig::NEXTCAR_TRC_images}(b)) over the reconstructed routes extracted from the central Ohio region. The results presented in this section are related to tests run over an urban route, as shown in Fig. \ref{fig::R_15_features}. Here, the reader is encouraged to refer to \cite{deshpande2021vehicle}, in which the in-vehicle implementation of the optimizer is compared and verified against the corresponding simulation models over a mixed (urban-highway) route.

For each of the tests conducted, the measured and calculated variables are vehicle speed, battery SoC, cumulative fuel consumed, and travel time. Each test case is repeated at least five times while ensuring that the controllable test conditions (such as the payload, initial battery SoC, calibration values for acceleration limits that affect comfort, wait time at a stop) along each trip are maintained similar. Factors that cannot be controlled (such as track conditions, ambient weather) are noted throughout the tests. This is used to evaluate the variability in travel time and fuel consumption for each case, and the possible inherent variability across fuel consumption measurements. In this study, the \underline{A}ir-\underline{F}uel \underline{R}atio (AFR) and airflow estimates from the ECM (obtained over CAN or Controller Area Network) are used to calculate the instantaneous fuel flow rate.

\subsubsection{Test Setup for Evaluation of Eco-Driving Algorithm}

For evaluation of the Eco-Driving case, three calibrations for the driver aggressiveness parameter are considered: $\gamma = \{0.3,0.7,0.75 \}$. In this scenario, all the traffic lights along the route are assumed to be flashing red, i.e. treated as stop signs. Table \ref{tab::test_setup_LO_RA} summarizes the test setup for the Eco-Driving test scenario. Note that the horizon length corresponding to $N_H = 20$ in this case takes a range of values, as the step size of the discretized distance grid is not fixed. This enables the usage of a fine discretization in certain sections of the route and coarser in other sections, allowing the Eco-Driving algorithm to satisfactorily capture the model dynamics while running in real-time on the MABx. 

\begin{table}[!t]
	\centering
	\begin{tabular}{cc}
		\hline
		Variable & Description \\ \hline
		HEV state  &  $\SI{48}{V}$ mHEV \\
		DSF state & ON \\
		Vehicle mass & $\approx \SI{1850}{kg}$ (incl. payload) \\
		Grade & $\SI{0}{\%}$ \\
		Initial SoC & $\SI{50}{\%}$  \\
		Controls & Eco-Driving Control \\
		Driver & ACC
	\end{tabular}

	\bigskip
	
	\begin{tabular}{cc}
		\hline
		Parameter & Value \\ \hline
		$N_H$ & $20$ ($150-750\si{m}$) \\
		$\gamma$ &  $\{0.3, 0.7, 0.75 \}$ \\
		$\{a^{min}, a^{max} \}$  & $\{-2.4, 2.4\}\si{m/s^2}$ \\
		$I_{bias}$ & $\SI{12}{A}$
	\end{tabular}
	\caption{Summary of Eco-Driving test setup and calibration parameters.}
	\label{tab::test_setup_LO_RA}
\end{table}

\subsubsection{Test Setup for Evaluation of Eco-Driving Algorithm with Eco-AND}

In order to evaluate the Eco-Driving algorithm with Eco-AND, a test environment was designed where traffic light SPaT information is broadcast to the optimizer as the vehicle approaches a signalized intersection. The introduction of SPaT variability makes evaluation of the Eco-Driving with Eco-AND case different from the aforementioned Eco-Driving case. For an ideal experimental verification of the Eco-Driving algorithm with Eco-AND, it would be necessary to run a statistically significant number of tests, treating $\gamma$ and SPaT as random variables. However, to significantly reduce the number of tests and the resulting testing time without compromising the validity of the results, sample cases from the Monte Carlo simulations (Fig. \ref{fig::MC_sim_results_R15}) were selected and reconstructed at TRC.

\begin{figure}[!t]
	\centering
	\vspace{-3mm}
	\subfloat[RSU apparatus]{\includegraphics[width=0.55\columnwidth]{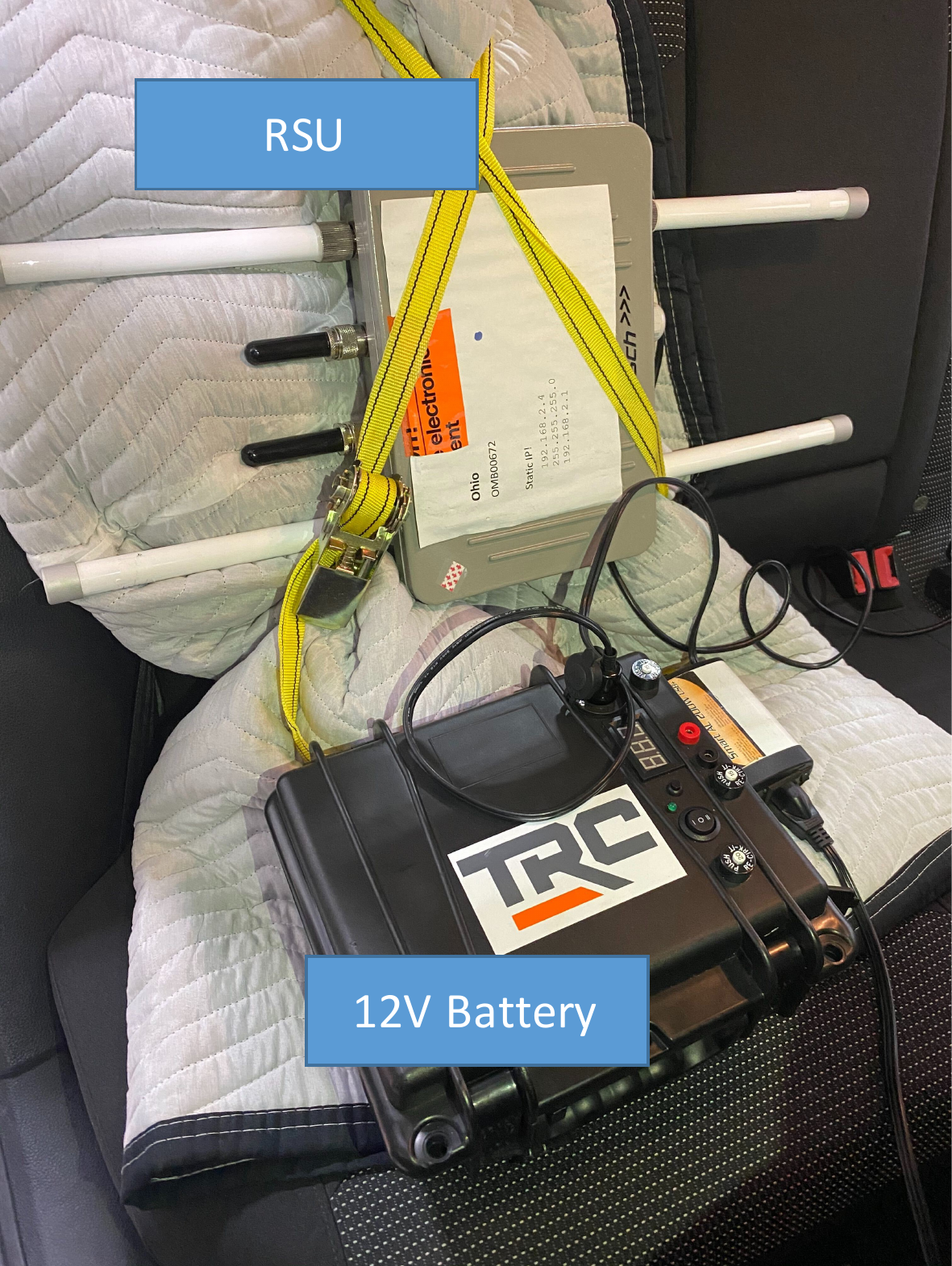}}
	\hfil
	\subfloat[Illustration of test vehicle approaching a signalized intersection]{\includegraphics[width=0.8\columnwidth]{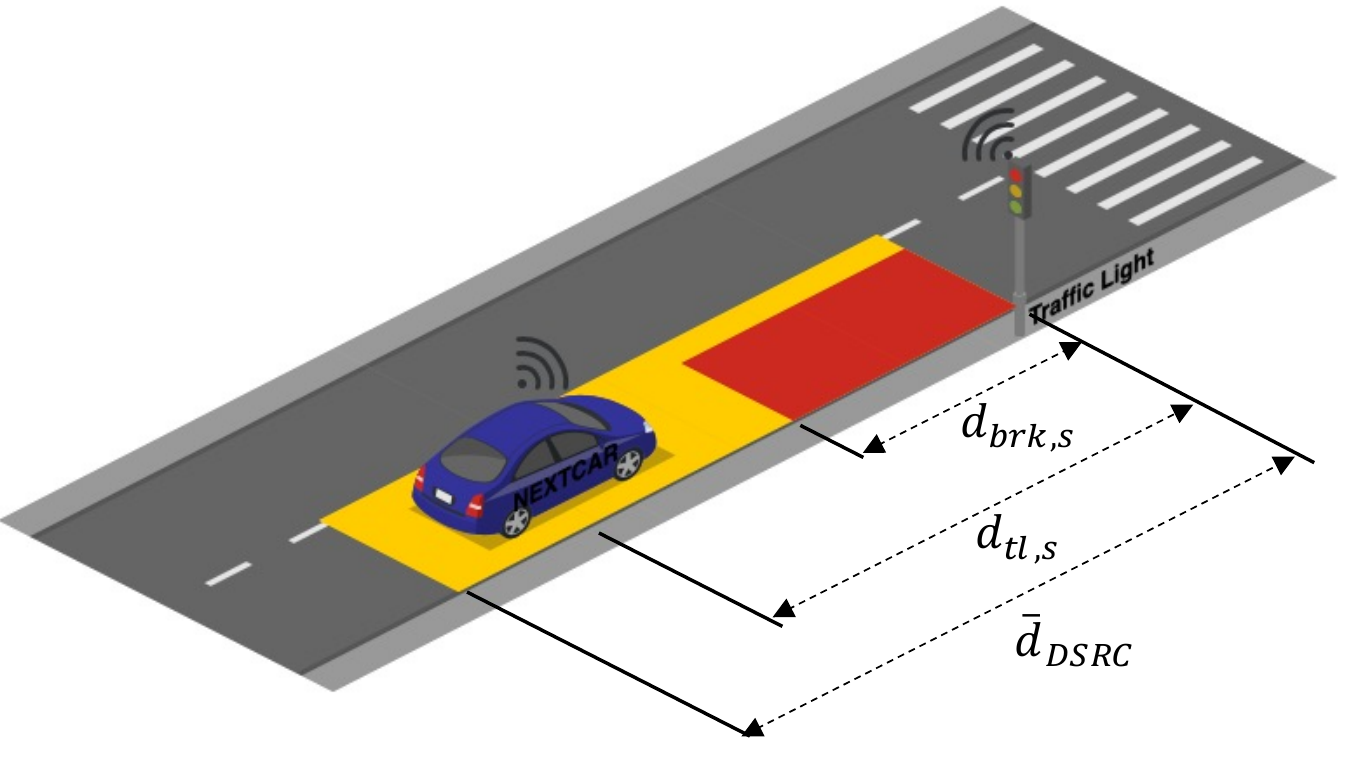}}
	\caption{RSU (mounted in the rear seat) that broadcasts SPaT information within DSRC range of the upcoming traffic light.}
	\label{fig::RSU_apparatus_PiGe_testing}
	\vspace{-1mm}
\end{figure}

For each tested condition, the values of $\gamma$ and the SPaT scenario were selected to represent a dominant mode of the respective Monte Carlo simulation over that route. In this work, $\gamma=0.7$ is chosen for the reconstructed route as it compactly represents normal driver aggressiveness. The SPaT scenarios are replicated in real-time by broadcasting them on a \underline{R}oad\underline{s}ide \underline{U}nit (RSU). For convenience in setup and testing, this RSU is mounted on the rear seat and connected to a supplementary $\SI{12}{V}$ battery as shown in Fig. \ref{fig::RSU_apparatus_PiGe_testing}. The V2I communication is emulated via the in-vehicle RSU that broadcasts SPaT and MAP information in accordance with the SAE J2735 standard to the \underline{O}n-\underline{B}oard \underline{U}nit (OBU) mounted on the vehicle.

A custom application was developed to input specific SPaT scenarios, defined by the duration of phase for each traffic light, and initial parameters corresponding to a particular SPaT scenario (i.e. initial phase and time remaining in initial phase). A custom decoder, designed and implemented on a PC connected to the OBU, then unpacks the SPaT messages and makes it available on the MABx, which contains the Eco-Driving algorithm with Eco-AND. Within the range $\bar{d}_{DSRC}$ from the upcoming traffic light, an arbitration strategy implemented within the MABx model selects the corresponding SPaT information.

Note that the travel time recorded for testing the Eco-Driving with Eco-AND case includes the time spent at stop signs and the wait time at red traffic lights along the route. Here, the test setup for evaluation is identical to that for the Eco-Driving case in Table \ref{tab::test_setup_LO_RA} but for the ``Controls" category, where the Eco-Driving algorithm equipped with Eco-AND capability is used.

%
%

\subsubsection{Test Setup for Baseline Evaluation}

The baseline considered for benchmarking the optimizer must be representative of the driving behavior of a conventional driver. However, utilizing human drivers over reconstructed routes limits the repeatability of the tests, inevitably introducing undesired variability in the fuel consumption and travel time.

For this reason, a BTR is programmed and calibrated to follow reference velocity profiles which are generated in simulation using the EDM calibrated to represent a variety of driving styles. This ensures repeatability and controlled variability across baseline tests. The BTR accurately tracks the reference velocity profile by robotic actuation of the accelerator and brake pedals with a human driver only responsible for the steering action. Table \ref{tab::test_setup_baseline} summarizes the test setup adopted to evaluate the baseline.

\begin{table}[!t]
	\centering
	\begin{tabular}{cc}
		\hline
		Variable & Description \\ \hline
		HEV state  &  $\SI{48}{V}$ mHEV \\
		DSF state & OFF \\
		Vehicle mass & $\approx \SI{1850}{kg}$ (incl. payload) \\
		Grade & $\SI{0}{\%}$ \\
		Initial SoC & $\SI{50}{\%}$  \\
		Controls & Production-level ECM \\
		Driver & BTR
	\end{tabular}
	\caption{Summary of baseline test setup.}
	\label{tab::test_setup_baseline}
\end{table}

Here, note that the processing of the baseline test results differ depending on whether the Eco-Driving or the Eco-Driving with Eco-AND is being benchmarked. For fair evaluation of the Eco-Driving case, the travel time recorded excludes the time spent at stops along the route, while for the Eco-Driving with Eco-AND case, the travel time includes the time spent at stop signs as well as the wait time at red traffic lights.

\subsection{Experimental Results for Eco-Driving}

To evaluate the benefits of the Eco-Driving algorithm over a representative baseline, multiple trips are executed over the same reconstructed urban route for the chosen $\gamma$ values (refer Table \ref{tab::test_setup_LO_RA}). The travel times from the Eco-Driving tests are used to generate the EDM parameter sets with similar (driver) aggressiveness and travel times for each $\gamma$. The generated EDM velocity profiles are then used as reference inputs to the BTR.

\begin{figure}[!t]
	\centering
	\vspace{-3mm}
	\subfloat[Pareto curve]{\includegraphics[width=0.8\columnwidth]{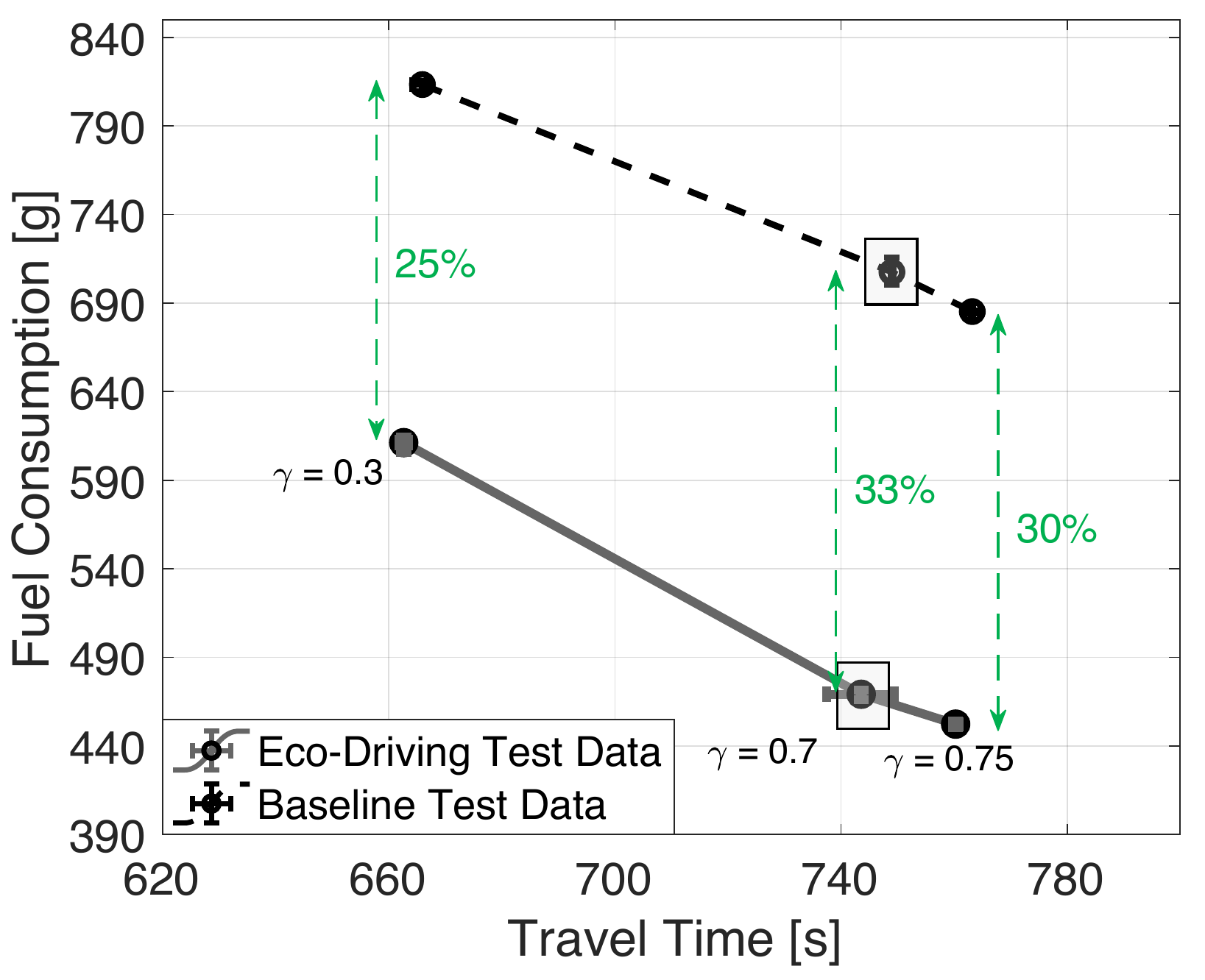}}
	\hfil
	\subfloat[Vehicle velocity and battery SoC profile, $\gamma = 0.7$]{\includegraphics[width=\columnwidth]{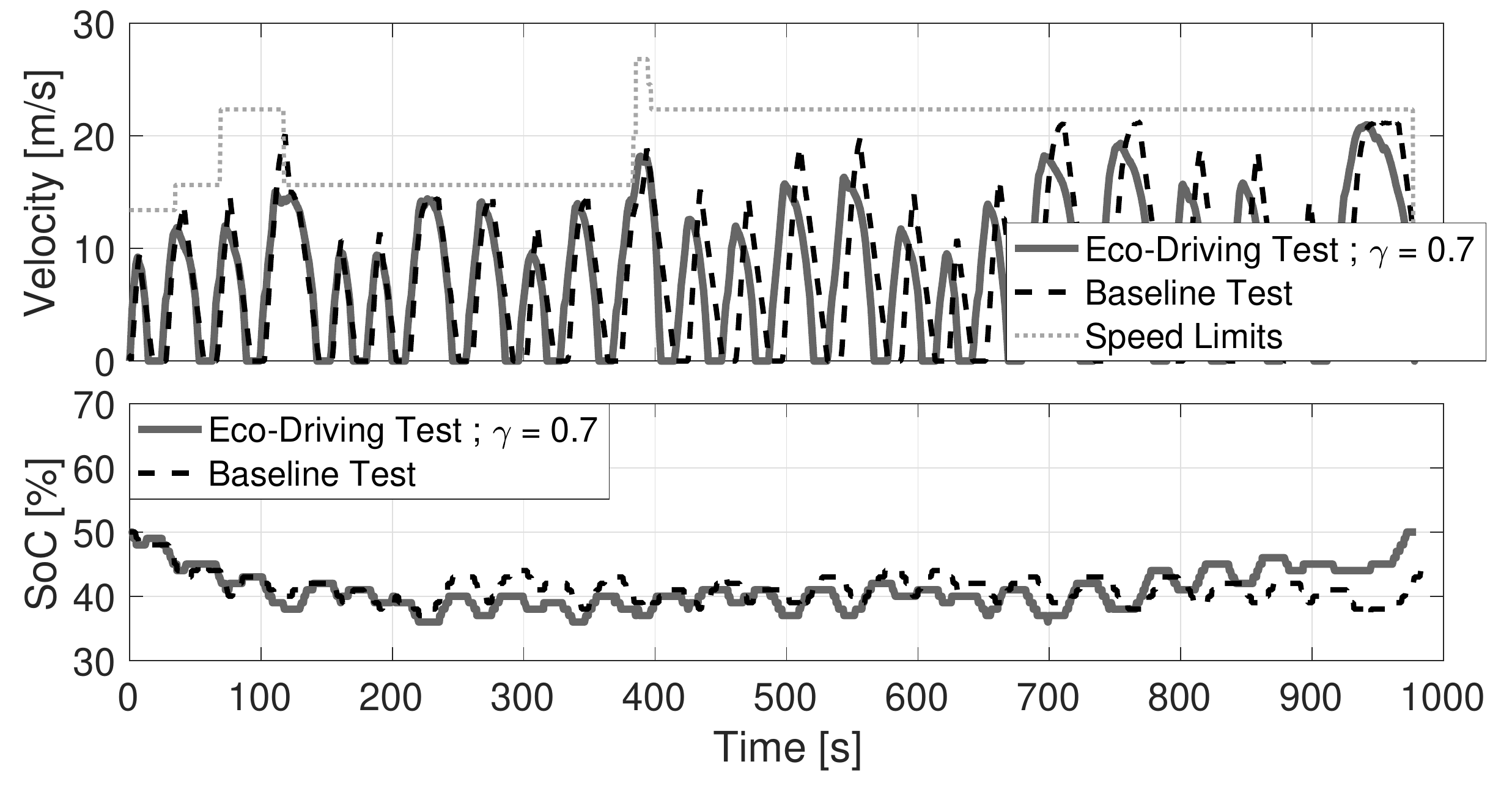}}
	\hfil
	\subfloat[Cumulative fuel consumption, $\gamma = 0.7$]{\includegraphics[width=0.8\columnwidth]{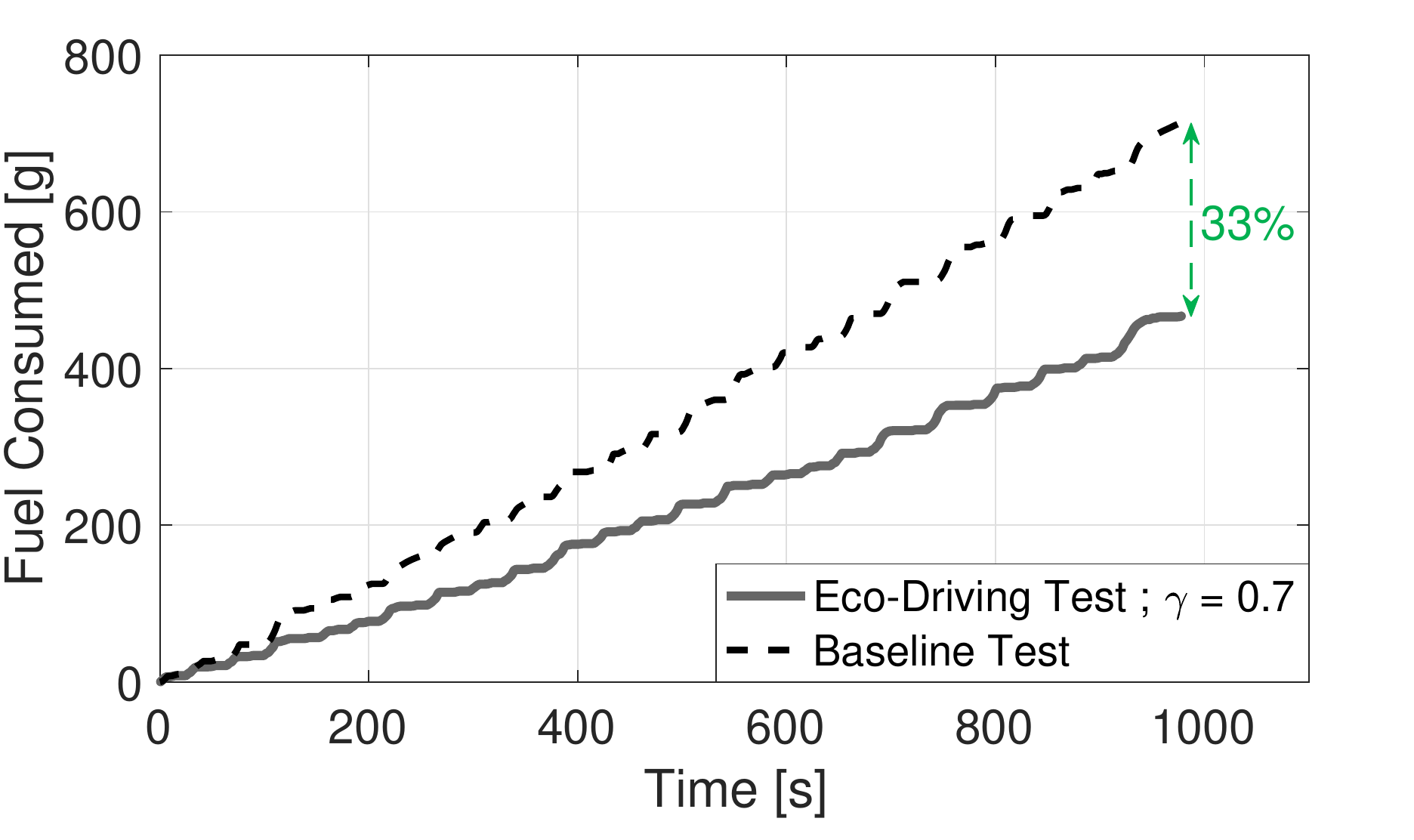}}
	\caption{In-vehicle testing of the Eco-Driving algorithm and comparison against baseline over urban route.}
	\label{fig::R15_test_baseline_opt_pareto_comp}
	\vspace{-1mm}
\end{figure}

Fig. \ref{fig::R15_test_baseline_opt_pareto_comp}(a) compares the vehicle testing results for the baseline (executed using the BTR) against the Eco-Driving case, for each $\gamma$. Here, the error bars represent the maximum deviation of the data points from their respective mean and are determined using data from at least five test runs. Over the urban route tested, featuring $23$ stops, the Eco-Driving algorithm with DSF enabled reduces the fuel consumption by over $\SI{25}{\%}$ (and up to $\SI{33}{\%}$), relative to the baseline, with comparable travel time.

In Fig. \ref{fig::R15_test_baseline_opt_pareto_comp}(b) and \ref{fig::R15_test_baseline_opt_pareto_comp}(c), a run-to-run comparison of the states (vehicle velocity and battery SoC) and cumulative fuel consumption along the urban route are respectively shown for $\gamma = 0.7$. A noticeable contrast is observed between the baseline and optimizer tests, with the Eco-Driving algorithm resulting in $\SI{33}{\%}$ fuel savings for similar travel time. Here, the DSF-enabled optimizer increases the operation of the engine within the DSF fly-zones and thereby leverages the fuel-saving potential of the cylinder deactivation technology (DSF). A notable result is that the Eco-Driving algorithm computes an energy management strategy that has a charge-sustaining SoC profile over the entire route (both in simulation and from in-vehicle implementation). This highlights the benefits of the closed-loop control strategy determined by the ADP-based optimizer and the inherent robustness it provides against modeling inaccuracies.

\subsection{Experimental Results for Eco-Driving with Eco-AND}

The specific case ($\gamma$ and SPaT scenario tuple) selected from the Monte Carlo simulations for vehicle testing corresponds to a dominant mode of the statistical distribution. Over the reconstructed urban route considered, the average (over five runs) in-vehicle fuel consumption and travel time are observed to lie within the range of their corresponding distributions. Further, as seen in Table \ref{tab::results_LO_RA_PiGe_test_sim_comp}, they are closely comparable to their respective mean values from the Monte Carlo simulations (refer Fig. \ref{fig::MC_sim_results_R15}).

\begin{figure}[!t]
	\centering
	\vspace{-3mm}
	\subfloat[Vehicle velocity, $\gamma = 0.7$]{\includegraphics[width=\columnwidth]{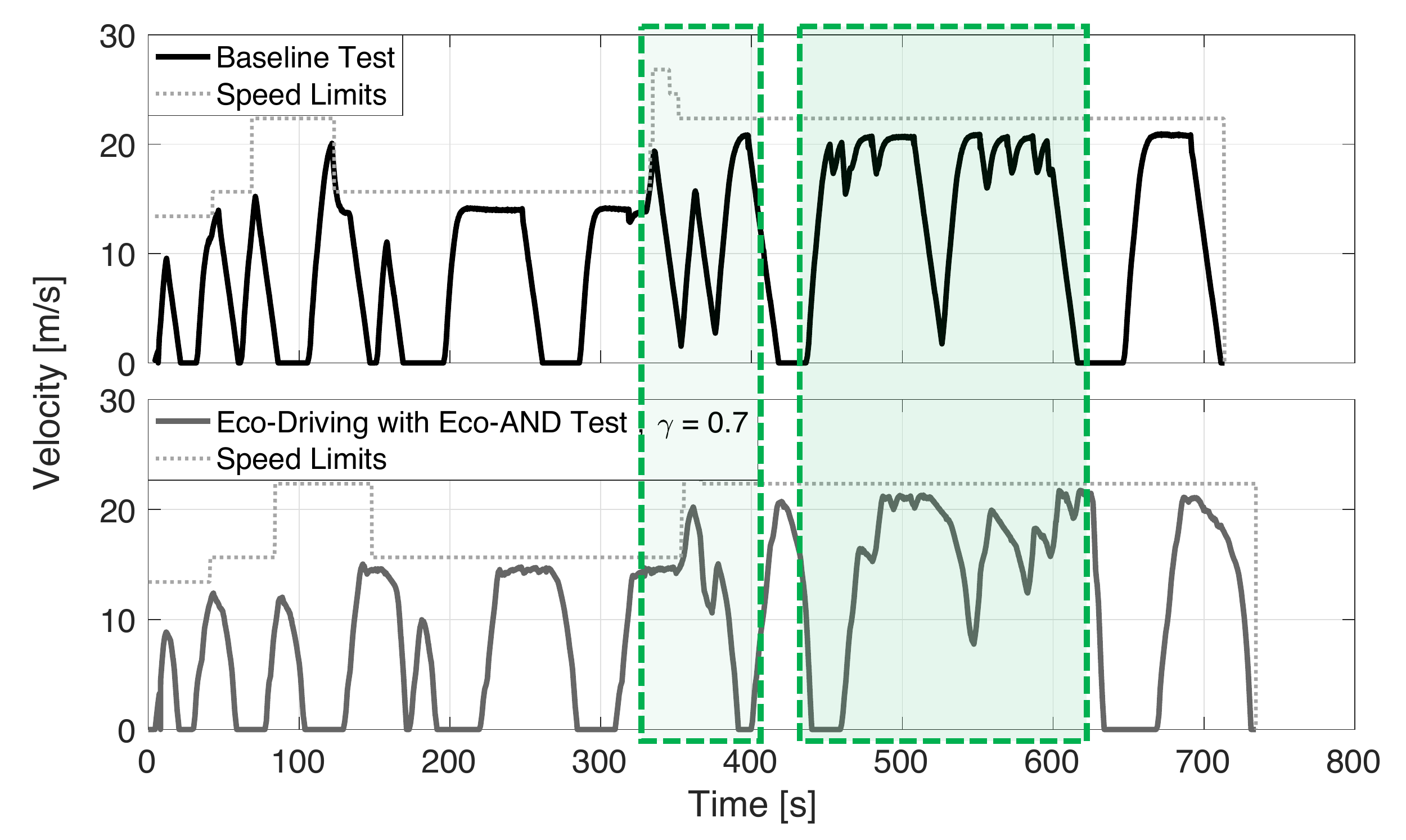}}
	\hfil
	\subfloat[Time-space plot, $\gamma = 0.7$]{\includegraphics[width=\columnwidth]{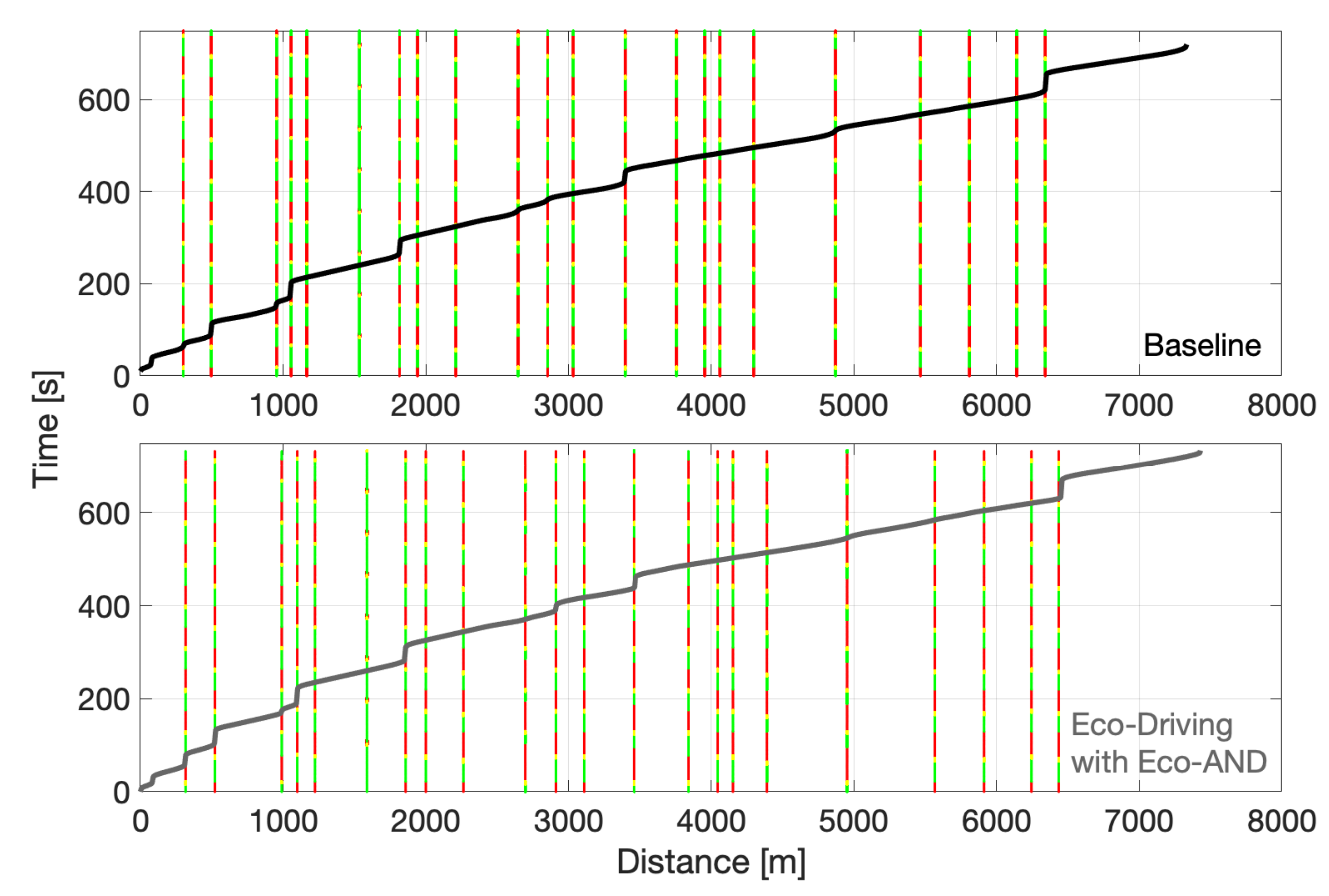}}
	\caption{Run-to-run comparison of vehicle velocity and time-space plots between baseline and Eco-Driving with Eco-AND experimental tests over urban route.}
	\label{fig::R15_test_baseline_opt_V_veh_tl_plot}
	\vspace{-1mm}
\end{figure}

The baseline case considered to benchmark the optimizer is constructed such that the driver aggressiveness (i.e., resulting travel time) matches a dominant mode of the statistical distribution obtained through Monte Carlo simulations while being comparable to the mean travel time from the Eco-Driving with Eco-AND testing. Fig. \ref{fig::R15_test_baseline_opt_V_veh_tl_plot} compares the in-vehicle test results from the baseline (using the BTR) and the Eco-Driving with Eco-AND cases. The enhanced range (compared to typical human LoS) and information (signal time in addition to the current phase) enabled by V2I technologies is used by the Eco-AND feature to provide modified speed constraints to the rollout algorithm-based optimization. While the number of stop-at-red scenarios remains the same in this specific case, the resulting velocity profile is much smoother with fewer acceleration-deceleration events (highlighted with the green shaded sections).


\begin{table}[!t]
	\centering
	\begin{tabular}{>{\centering \arraybackslash}m{0.2\columnwidth}>{\centering \arraybackslash}m{0.25\columnwidth}>{\centering \arraybackslash}m{0.3\columnwidth}}
		\hline
		Case & Travel Time [\si{s}] & Fuel Consumed [\si{g}] \\ \hline
		Virtual (Monte Carlo simulation)& $712$  & $404$ \\
		Experimental tests & $735$ & $417$
	\end{tabular}
	\caption{Mean travel time and fuel consumption results (virtual and experimental) using LO-RA with Eco-AND over urban route, $\gamma = 0.7$.}
	\label{tab::results_LO_RA_PiGe_test_sim_comp}
\end{table}

\begin{figure}[!t]
	\centering
	\vspace{-3mm}
	\subfloat[Battery SoC, $\gamma = 0.7$]{\includegraphics[width=0.8\columnwidth]{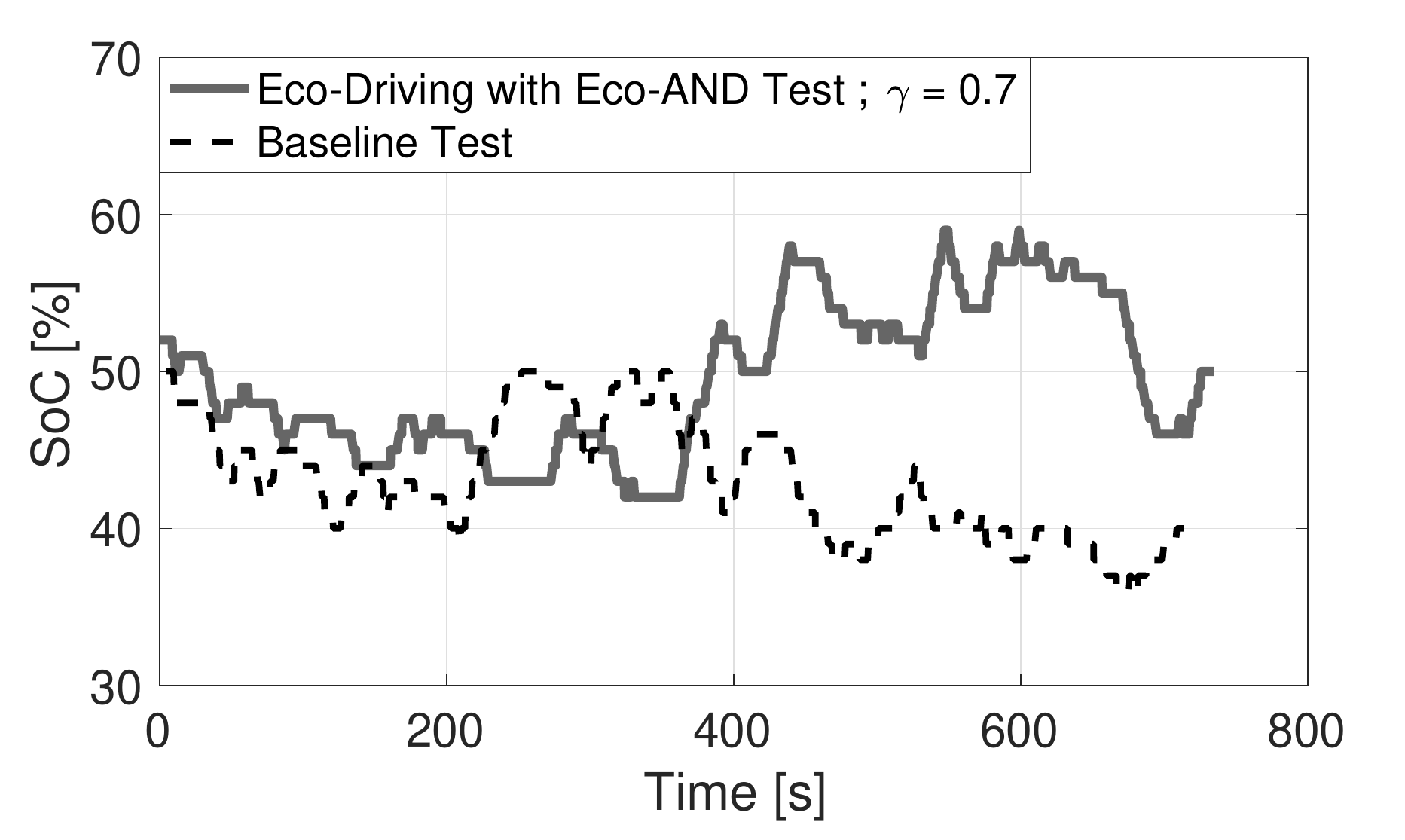}}
	\hfil
	\subfloat[Cumulative fuel consumption, $\gamma = 0.7$]{\includegraphics[width=0.8\columnwidth]{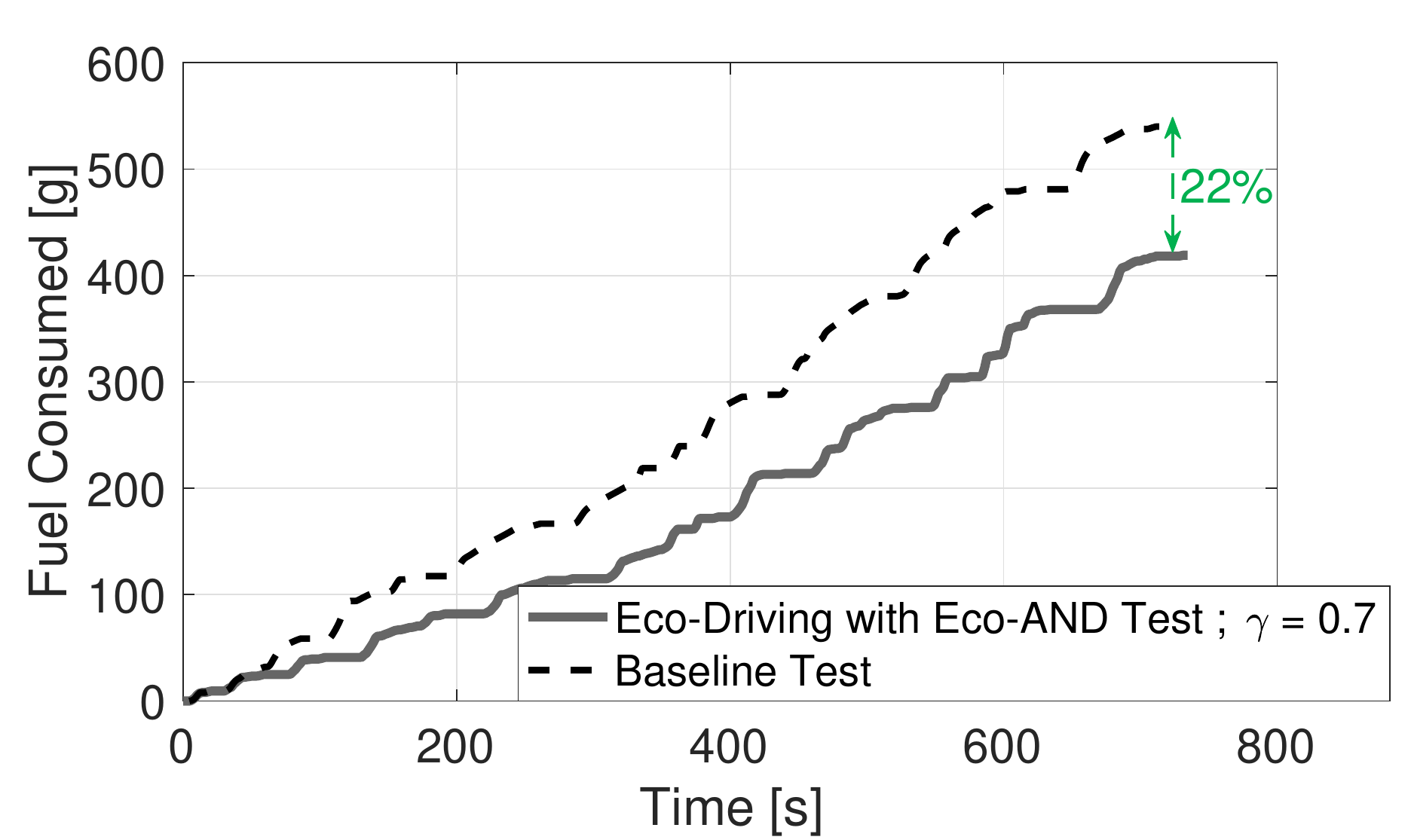}}
	\caption{Run-to-run comparison of battery SoC and cumulative fuel consumption profiles between baseline and Eco-Driving with Eco-AND experimental tests over urban route.}
	\label{fig::R15_test_baseline_opt_SoC_fuel}
	\vspace{-1mm}
\end{figure}

The torque split strategies of the baseline and Eco-Driving with Eco-AND cases over the urban route are evaluated by comparing the resulting battery SoC and cumulative fuel consumption trajectories, shown in Fig. \ref{fig::R15_test_baseline_opt_SoC_fuel}. A key benefit from the designed Eco-Driving algorithm with Eco-AND is the resulting charge sustaining torque split strategy. In contrast, the terminal SoC for the baseline case is $\SI{40}{\%}$ (while its initial SoC was $\SI{50}{\%}$).

\begin{table}[!t]
	\centering
	\begin{tabular}{>{\centering \arraybackslash}m{0.25\columnwidth}>{\centering \arraybackslash}m{0.25\columnwidth}>{\centering \arraybackslash}m{0.3\columnwidth}}
		\hline
		Case & Travel Time [\si{s}] & Fuel Consumed [\si{g}] \\ \hline
		Baseline & $714$  & $538$ \\
		Eco-Driving with Eco-AND & $735$ & $417$
	\end{tabular}
	\caption{Comparison of cumulative fuel consumption and travel time from baseline (without DSF) and Eco-Driving with Eco-AND (with DSF) experimental tests.}
	\label{tab::results_LO_RA_PiGe_test_baseline_comp}
\end{table}

The impact of smoother velocity transients, and fuel-efficient approach and departure from signalized intersections and stop signs is seen in the relative cumulative fuel consumption. For the specific route tested here, the optimizer saves over $\SI{22}{\%}$ in fuel with a marginal increase of $\SI{2.9}{\%}$ in trip time while ensuring SoC neutrality over the entire trip.

\section{Conclusions}

In this paper, a multi-layer hierarchical MPC framework is developed to solve the Eco-Driving problem for a $\SI{48}{V}$ mild-hybrid powertrain in a connected vehicle environment. The proposed control framework comprises a long-term co-optimization of vehicle speed and SoC performed over the entire route itinerary, and a short-term optimization where the variability in route conditions and/or uncertainty in route information are considered by transforming the full-route optimization into a receding horizon optimal control problem, solved periodically over shorter horizons. The terminal cost for the receding horizon optimization is approximated as the residual cost (or cost to complete the remaining route) from the full-route optimization and solved using Approximate Dynamic Programming, specifically the rollout algorithm. This approximation, termed as the base policy in this paper, results in cost improvement, which is important for real-time applications with environmental and modeling uncertainties.

Further, this work proposes a systematic and comprehensive verification framework for virtual and experimental evaluation (on a test vehicle equipped with a rapid prototyping embedded controller). Here, the fuel-saving potential of the hierarchical optimization framework was evaluated over real-world routes. Results indicate that both the simulation and track testing show a fuel consumption reduction of more than $\SI{20}{\%}$, compared against a suitable baseline with comparable trip time. Further, the simulation and track testing results are sufficiently close to each other, validating that the developed rollout algorithm-based optimization framework is suitable for real-time applications containing environmental disturbances and modeling uncertainties. Future work includes the implementation of the algorithm with the consideration of traffic information via vehicle-to-vehicle (V2V) and signal phase and timing information via vehicle-to-infrastructure (V2I) communication.

\begin{acknowledgment}
	
This paper is supported by the United States Department of Energy, Advanced Research Projects Agency -- Energy (Award Number DE-AR0000794). The authors gratefully acknowledge BorgWarner Inc. (formerly Delphi Technologies) for providing continued technical support and for the insightful discussions.

\end{acknowledgment}

%
% ...

\bibliographystyle{asmems4}

\bibliography{references}

\begin{thebibliography}{10}

\bibitem{guanetti2018control}
Guanetti, J., Kim, Y., and Borrelli, F., 2018.
\newblock ``Control of connected and automated vehicles: State of the art and
  future challenges''.
\newblock {\em Annual reviews in control, {\bf 45}}, pp.~18--40.

\bibitem{gupta2020estimation}
Gupta, S., Deshpande, S.~R., Tufano, D., Canova, M., Rizzoni, G., Aggoune, K.,
  Olin, P., and Kirwan, J., 2020.
\newblock Estimation of fuel economy on real-world routes for next-generation
  connected and automated hybrid powertrains.
\newblock Tech. rep., SAE Technical Paper.

\bibitem{xu2018design}
Xu, S., and Peng, H., 2018.
\newblock ``Design and comparison of fuel-saving speed planning algorithms for
  automated vehicles''.
\newblock {\em IEEE Access, {\bf 6}}, pp.~9070--9080.

\bibitem{guzzella2007vehicle}
Guzzella, L., Sciarretta, A., et~al., 2007.
\newblock {\em Vehicle propulsion systems}, Vol.~1.
\newblock Springer.

\bibitem{alam2014critical}
Alam, M.~S., and McNabola, A., 2014.
\newblock ``A critical review and assessment of eco-driving policy \&
  technology: Benefits \& limitations''.
\newblock {\em Transport Policy, {\bf 35}}, pp.~42--49.

\bibitem{sciarretta2015optimal}
Sciarretta, A., De~Nunzio, G., and Ojeda, L.~L., 2015.
\newblock ``Optimal ecodriving control: Energy-efficient driving of road
  vehicles as an optimal control problem''.
\newblock {\em IEEE Control Systems Magazine, {\bf 35}}(5), pp.~71--90.

\bibitem{jin2016power}
Jin, Q., Wu, G., Boriboonsomsin, K., and Barth, M.~J., 2016.
\newblock ``Power-based optimal longitudinal control for a connected
  eco-driving system''.
\newblock {\em IEEE Transactions on Intelligent Transportation Systems, {\bf
  17}}(10), pp.~2900--2910.

\bibitem{ozatay2014cloud}
Ozatay, E., Onori, S., Wollaeger, J., Ozguner, U., Rizzoni, G., Filev, D.,
  Michelini, J., and Di~Cairano, S., 2014.
\newblock ``Cloud-based velocity profile optimization for everyday driving: A
  dynamic-programming-based solution''.
\newblock {\em IEEE Transactions on Intelligent Transportation Systems, {\bf
  15}}(6), pp.~2491--2505.

\bibitem{han2019fundamentals}
Han, J., Vahidi, A., and Sciarretta, A., 2019.
\newblock ``Fundamentals of energy efficient driving for combustion engine and
  electric vehicles: An optimal control perspective''.
\newblock {\em Automatica, {\bf 103}}, pp.~558--572.

\bibitem{mensing2012vehicle}
Mensing, F., Trigui, R., and Bideaux, E., 2012.
\newblock ``Vehicle trajectory optimization for hybrid vehicles taking into
  account battery state-of-charge''.
\newblock In 2012 IEEE vehicle power and propulsion conference, IEEE,
  pp.~950--955.

\bibitem{guo2016optimal}
Guo, L., Gao, B., Gao, Y., and Chen, H., 2016.
\newblock ``Optimal energy management for hevs in eco-driving applications
  using bi-level mpc''.
\newblock {\em IEEE Transactions on Intelligent Transportation Systems, {\bf
  18}}(8), pp.~2153--2162.

\bibitem{amini2019sequential}
Amini, M.~R., Gong, X., Feng, Y., Wang, H., Kolmanovsky, I., and Sun, J., 2019.
\newblock ``Sequential optimization of speed, thermal load, and power split in
  connected hevs''.
\newblock In 2019 American Control Conference (ACC), IEEE, pp.~4614--4620.

\bibitem{hao2018eco}
Hao, P., Wu, G., Boriboonsomsin, K., and Barth, M.~J., 2018.
\newblock ``Eco-approach and departure (ead) application for actuated signals
  in real-world traffic''.
\newblock {\em IEEE Transactions on Intelligent Transportation Systems, {\bf
  20}}(1), pp.~30--40.

\bibitem{ye2018prediction}
Ye, F., Hao, P., Qi, X., Wu, G., Boriboonsomsin, K., and Barth, M.~J., 2018.
\newblock ``Prediction-based eco-approach and departure at signalized
  intersections with speed forecasting on preceding vehicles''.
\newblock {\em IEEE Transactions on Intelligent Transportation Systems, {\bf
  20}}(4), pp.~1378--1389.

\bibitem{altan2017glidepath}
Altan, O.~D., Wu, G., Barth, M.~J., Boriboonsomsin, K., and Stark, J.~A., 2017.
\newblock ``Glidepath: Eco-friendly automated approach and departure at
  signalized intersections''.
\newblock {\em IEEE Transactions on Intelligent Vehicles, {\bf 2}}(4),
  pp.~266--277.

\bibitem{sun2018robust}
Sun, C., Guanetti, J., Borrelli, F., and Moura, S., 2018.
\newblock ``Robust eco-driving control of autonomous vehicles connected to
  traffic lights''.
\newblock {\em arXiv preprint arXiv:1802.05815}.

\bibitem{qin2003survey}
Qin, S.~J., and Badgwell, T.~A., 2003.
\newblock ``A survey of industrial model predictive control technology''.
\newblock {\em Control engineering practice, {\bf 11}}(7), pp.~733--764.

\bibitem{borrelli2017predictive}
Borrelli, F., Bemporad, A., and Morari, M., 2017.
\newblock {\em Predictive control for linear and hybrid systems}.
\newblock Cambridge University Press.

\bibitem{bellman1966dynamic}
Bellman, R., 1966.
\newblock ``Dynamic programming''.
\newblock {\em Science, {\bf 153}}(3731), pp.~34--37.

\bibitem{lee2010approximate}
Lee, J.~H., and Wong, W., 2010.
\newblock ``Approximate dynamic programming approach for process control''.
\newblock {\em Journal of Process Control, {\bf 20}}(9), pp.~1038--1048.

\bibitem{yin2016energy}
Yin, J., Tang, T., Yang, L., Gao, Z., and Ran, B., 2016.
\newblock ``Energy-efficient metro train rescheduling with uncertain
  time-variant passenger demands: An approximate dynamic programming
  approach''.
\newblock {\em Transportation Research Part B: Methodological, {\bf 91}},
  pp.~178--210.

\bibitem{wilcutts2013design}
Wilcutts, M., Switkes, J., Shost, M., and Tripathi, A., 2013.
\newblock ``Design and benefits of dynamic skip fire strategies for cylinder
  deactivated engines''.
\newblock {\em SAE International Journal of Engines, {\bf 6}}(1), pp.~278--288.

\bibitem{wilcutts2018electrified}
Wilcutts, M., Nagashima, M., Eisazadeh-Far, K., Younkins, M., and Confer, K.,
  2018.
\newblock Electrified dynamic skip fire (edsf): Design and benefits.
\newblock Tech. rep., SAE Technical Paper.

\bibitem{zhu2021gpu}
Zhu, Z., Gupta, S., Pivaro, N., Deshpande, S.~R., and Canova, M., 2021.
\newblock ``A gpu implementation of a look-ahead optimal controller for
  eco-driving based on dynamic programming''.
\newblock {\em arXiv preprint arXiv:2104.01284}.

\bibitem{bertsekas1995dynamic}
Bertsekas, D.~P., 1995.
\newblock {\em Dynamic programming and optimal control}, Vol.~1.
\newblock Athena scientific Belmont, MA.

\bibitem{johannesson2008approximate}
Johannesson, L., and Egardt, B., 2008.
\newblock ``Approximate dynamic programming applied to parallel hybrid
  powertrains''.
\newblock {\em IFAC proceedings volumes, {\bf 41}}(2), pp.~3374--3379.

\bibitem{hellstrom2010design}
Hellstr{\"o}m, E., {\AA}slund, J., and Nielsen, L., 2010.
\newblock ``Design of an efficient algorithm for fuel-optimal look-ahead
  control''.
\newblock {\em Control Engineering Practice, {\bf 18}}(11), pp.~1318--1327.

\bibitem{bae2019real}
Bae, S., Choi, Y., Kim, Y., Guanetti, J., Borrelli, F., and Moura, S., 2019.
\newblock ``Real-time ecological velocity planning for plug-in hybrid vehicles
  with partial communication to traffic lights''.
\newblock In 2019 IEEE 58th Conference on Decision and Control (CDC), IEEE,
  pp.~1279--1285.

\bibitem{bertsekas1996neuro}
Bertsekas, D.~P., and Tsitsiklis, J.~N., 1996.
\newblock {\em Neuro-dynamic programming}.
\newblock Athena Scientific.

\bibitem{melo2008analysis}
Melo, F.~S., Meyn, S.~P., and Ribeiro, M.~I., 2008.
\newblock ``An analysis of reinforcement learning with function
  approximation''.
\newblock In Proceedings of the 25th international conference on Machine
  learning, pp.~664--671.

\bibitem{zhu2020energy}
Zhu, Z., Liu, Y., and Canova, M., 2020.
\newblock ``Energy management of hybrid electric vehicles via deep
  q-networks''.
\newblock In 2020 American Control Conference (ACC), IEEE, pp.~3077--3082.

\bibitem{bertsekas2005rollout}
Bertsekas, D., 2005.
\newblock ``Rollout algorithms for constrained dynamic programming''.
\newblock {\em Lab. for Information and Decision Systems Report, {\bf 2646}}.

\bibitem{sun2020optimal}
Sun, C., Guanetti, J., Borrelli, F., and Moura, S.~J., 2020.
\newblock ``Optimal eco-driving control of connected and autonomous vehicles
  through signalized intersections''.
\newblock {\em IEEE Internet of Things Journal, {\bf 7}}(5), pp.~3759--3773.

\bibitem{larson1978principles}
Larson, R.~E., Casti, J.~L., and Casti, J.~L., 1978.
\newblock {\em Principles of dynamic programming}, Vol.~7.
\newblock M. Dekker New York.

\bibitem{gupta2019enhanced}
Gupta, S., Deshpande, S.~R., Tulpule, P., Canova, M., and Rizzoni, G., 2019.
\newblock ``An enhanced driver model for evaluating fuel economy on real-world
  routes''.
\newblock {\em IFAC-PapersOnLine, {\bf 52}}(5), pp.~574--579.

\bibitem{coovert2009design}
Coovert, D.~A., Heydinger, G.~J., Bixel, R.~A., Andreatta, D., Guenther, D.~A.,
  Sidhu, A.~S., and Mikesell, D.~R., 2009.
\newblock ``Design and operation of a brake and throttle robot''.
\newblock {\em SAE International Journal of Passenger Cars-Mechanical Systems,
  {\bf 2}}(2009-01-0429), pp.~613--621.

\bibitem{rajakumar2020benchmarking}
Rajakumar~Deshpande, S., Gupta, S., Kibalama, D., Pivaro, N., and Canova, M.,
  2020.
\newblock ``Benchmarking fuel economy of connected and automated vehicles in
  real world driving conditions via monte carlo simulation''.
\newblock In Dynamic Systems and Control Conference, Vol.~84270, American
  Society of Mechanical Engineers, p.~V001T10A004.

\bibitem{fathy2008online}
Fathy, H.~K., Kang, D., and Stein, J.~L., 2008.
\newblock ``Online vehicle mass estimation using recursive least squares and
  supervisory data extraction''.
\newblock In 2008 American control conference, IEEE, pp.~1842--1848.

\bibitem{vahidi2005recursive}
Vahidi, A., Stefanopoulou, A., and Peng, H., 2005.
\newblock ``Recursive least squares with forgetting for online estimation of
  vehicle mass and road grade: theory and experiments''.
\newblock {\em Vehicle System Dynamics, {\bf 43}}(1), pp.~31--55.

\bibitem{behrisch2011sumo}
Behrisch, M., Bieker, L., Erdmann, J., and Krajzewicz, D., 2011.
\newblock ``Sumo--simulation of urban mobility: an overview''.
\newblock In Proceedings of SIMUL 2011, The Third International Conference on
  Advances in System Simulation, ThinkMind.

\bibitem{krajzewicz2002sumo}
Krajzewicz, D., Hertkorn, G., R{\"o}ssel, C., and Wagner, P., 2002.
\newblock ``Sumo (simulation of urban mobility)-an open-source traffic
  simulation''.
\newblock In Proceedings of the 4th middle East Symposium on Simulation and
  Modelling (MESM20002), pp.~183--187.

\bibitem{citycbus2021tsdm}
{The City of Columbus}, 2021.
\newblock Traffic signal design manual.
\newblock
  \url{https://www.columbus.gov/publicservice/Design-and-Construction/document-library/Traffic-Signal-Design-Manual/}.

\bibitem{deshpande2021vehicle}
Deshpande, S.~R., Gupta, S., Kibalama, D., Pivaro, N., Canova, M., Rizzoni, G.,
  Aggoune, K., Olin, P., and Kirwan, J., 2021.
\newblock In-vehicle test results for advanced propulsion and vehicle system
  controls using connected and automated vehicle information.
\newblock Tech. rep., SAE Technical Paper.

\end{thebibliography}

\appendix       
\section{Proof of Cost Improvement in Rollout}
\label{app:cost_improvement_proof}
This proof is adapted from \cite{bertsekas1995dynamic} to better represent the nomenclature used in this work. Let $\tilde{J}_k(x_k)$ and $\hat{J}_k(x_k)$ be the costs-to-go of the base policy and rollout policy respectively, starting from a state $x_k$ at position $k, \quad \forall k = 1, \dots, N$. Define $\mathcal{M} := \left(\mu_{1}, \dots, \mu_{N} \right)$ as the base policy and $\mathcal{\hat{M}} := \left(\hat{\mu}_{1}, \dots, \hat{\mu}_{N} \right)$ as the rollout policy. At the terminal stage, let $\hat{J}_{N+1}(x_{N+1}) = \tilde{J}_{N+1}(x_{N+1}) = c_{N+1}(x_{N+1})$. The induction begins by assuming: $\hat{J}_{k+1}(x_{k+1}) \leq \tilde{J}_{k+1}(x_{k+1}), \quad \forall x_{k+1}$. Now, for all $x_k$:
\begin{equation}
\label{eq::cost_improvement_rollout}
\begin{aligned}
\hat{J}_{k}(x_{k}) &= \hat{J}_{k+1} \left(f_k(x_k, \hat{\mu}_k^*(x_k)) \right) + c_k(x_k, \hat{\mu}_k^*(x_k)) \\
& \leq \tilde{J}_{k+1} \left(f_k(x_k, \hat{\mu}_k^*(x_k)) \right) + c_k(x_k, \hat{\mu}_k^*(x_k)) \\
& \leq \tilde{J}_{k+1} \left(f_k(x_k, \mu_k^* (x_k)) \right) + c_k(x_k, \mu_k^*(x_k)) \\
&= \tilde{J}_k
\end{aligned}
\end{equation}
The proof is hence complete.


\end{document}